\documentclass[aps, prx, reprint, amsmath, amssymb, superscriptaddress, showpacs, floatfix, longbibliography]{revtex4-2}

\usepackage[dvipsnames,table]{xcolor}
\definecolor{linkcolor}{rgb}{0.16, 0.32, 0.75}
\definecolor{citecolor}{rgb}{0.0, 0.5, 0.0}
\definecolor{urlcolor}{rgb}{0.06, 0.46, 1.0}
\usepackage[colorlinks, linkcolor={linkcolor}, citecolor={citecolor}, urlcolor={urlcolor}]{hyperref}
\usepackage{bookmark}
\usepackage{mathtools}
\usepackage{tabularray}
\usepackage{physics}
\usepackage{mleftright}
\mleftright

\renewenvironment{pmatrix}{\left(\begin{tblr}{
  columns={c,colsep=2pt,mode=dmath},
  row{1}={abovesep=0pt},
  row{Z}={belowsep=0pt},
  column{1}={leftsep=0pt},
  column{Z}={rightsep=0pt}
}}{\end{tblr}\right)}

\begin{document}

\title{Density of scattering resonances in a disordered system}

\author{M.\,S.\,Kurilov}
\affiliation{Nanocenter CENN, SI-1000 Ljubljana, Slovenia}
\affiliation{Department of Physics, University of Ljubljana, SI-1000 Ljubljana, Slovenia}

\author{P.\,M.\,Ostrovsky}
\affiliation{Max Planck Institute for Solid State Research, 70569 Stuttgart, Germany}

\begin{abstract}
Reflection of particles from a disordered or chaotic medium is characterized by a scattering matrix that can be represented as a superposition of resonances. Each resonance corresponds to an eigenstate inside the medium and has a width related to the decay time of this eigenstate. We develop a general approach to study the distribution function of these resonance widths based on the nonlinear sigma model. We derive an integral representation of the distribution function that works equally well for systems of any symmetry and for any type of coupling to the measuring device. From this integral representation we find explicit analytic expressions for the distribution function in the case of disordered metallic grains. We also compare the analytic results to large-scale numerical simulations and observe their perfect agreement.
\end{abstract}

\maketitle

\tableofcontents

\section{Introduction}

Statistical properties of quantum scattering in disordered or chaotic systems play a major role in understanding diverse physical phenomena ranging from electron transport in metals to light propagation in optical fibers. One particular such statistical characteristic is the distribution of poles of the scattering matrix in the complex plane of energies. Over the past few decades, this distribution is actively studied both theoretically \cite{Sokolov1989, Haake1992, Fyodorov1997, Fyodorov1999, Sommers1999, Gluck2002, Fyodorov2012, Fyodorov2024, Fyodorov2025} and experimentally \cite{Kuhl2008, Falco2012, Liu2014} for setups of various physical origin, symmetry, and geometry. Imaginary parts of the scattering poles are particularly important since they are directly related to the widths of metastable quantum states (resonances) in a sample and determine Wigner scattering time delays in the system \cite{Chen2021}.

There are two alternative physical setups to study quantum scattering statistics involving either chaotic or disordered systems. In the former case, a particular microscopic Hamiltonian is chosen usually in the form of a quantum billiard or a set of regular scattering obstacles. When such a Hamiltonian exhibits classical chaotic dynamics, it can be studied semiclassically to infer statistics of quantum scattering \cite{Stoeckmann1999}. Semiclassical approximation assumes the scattering particle wave length to be much shorter than any geometrical size of the considered system. An alternative setup implies a disordered system described by a distribution of Hamiltonians with some randomness. Scattering properties are then averaged over this distribution. This stochastic approach also relies on the semiclassical approximation implying that the scattering particle wave length is much shorter than the average mean free path in the disordered system.

Relatively small disordered systems can be efficiently described within the random matrix theory (RMT) formalism of Wigner and Dyson \cite{Mehta2004}. The Hamiltonian of a disordered sample is replaced by a random Hermitian matrix with all matrix elements identically and independently distributed according to the Gaussian law. The only relevant information about the disordered system retained by RMT is the overall symmetry. Three symmetry classes are distinguished in the Wigner-Dyson RMT: unitary class (time-reversal symmetry broken, random complex Hermitian Hamiltonian), orthogonal class (time-reversal and spin symmetries preserved, random real symmetric Hamiltonian), and symplectic class (time-reversal preserved but spin symmetry broken, random Hamiltonian composed of real quaternions). Since microscopic features of the system are disregarded in this approach, resulting statistical characteristics appear to be universal.

RMT equally well describes statistical properties of both quantum chaotic and disordered systems. In the case of quantum chaos, this statement is known as the Bohigas-Giannoni-Schmidt conjecture \cite{Bohigas1984}. Although not formally proven, the conjecture is supported by an overwhelming amount of numerical evidence \cite{Wintgen1986, Delande1986, Sieber1990}. Applicability of RMT, and hence universality of the results, relies on the fact that wave functions extend to the whole volume of the sample. This is true at energy scales smaller than the Thouless energy $E_\text{Th} = \hbar/\tau_\text{erg}$ where $\tau_\text{erg}$ is the typical time needed for a particle to explore the full sample volume. The RMT approach is also very convenient for numerical studies of scattering \cite{Fyodorov2015, Haake1992}.

A microscopic proof of the universality conjecture for disordered metals (at a physical level of rigor) was obtained by Efetov \cite{Efetov1982}. It was achieved by developing an effective field theory for disordered systems---nonlinear supersymmetric sigma model \cite{Efetov1996}. This model was originally derived to study disordered mesoscopic metallic systems, hence we will use the terminology from this field in our analysis. The nonlinear sigma model is more general than RMT and is capable to describe extended systems whose size may exceed the typical diffusion distance at a given energy. The only assumption, aside from the semiclassical condition discussed above, is the ``dirty'' limit implying that the mean free path is shorter than any characteristic dimensions of the sample. For relatively small systems at energies below $E_\text{Th}$, the nonlinear sigma model reproduces the results of RMT \cite{Efetov1982}. The latter corresponds to the zero-dimensional limit of the sigma model. In extended samples, when the wave functions do not spread over the entire volume, Anderson localization effects become important. The nonlinear sigma model also provides a very efficient tool to study these phenomena \cite{Evers2008}.

\begin{figure}
\includegraphics[width=0.45\textwidth]{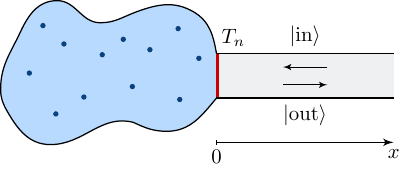}
\caption{Schematic depiction of a disordered sample attached to a ballistic measuring probe with $M$ conducting channels. The two parts of the system are coupled through a generic barrier characterized by a set of transmission probabilities $T_n$ in individual channels. Amplitudes of incident and reflected waves in the measuring lead are denoted by $\ket{\text{in}}$ and $\ket{\text{out}}$.}
\label{Fig:system}
\end{figure}

In the present paper, we will consider a general disordered system of arbitrary shape coupled to a relatively narrow ballistic lead, see Fig.\ \ref{Fig:system}. This measuring lead is used to probe particle scattering off the sample. Very generally, the scattering is described by the energy-dependent unitary scattering matrix $S_E$ whose size equals the number of conducting channels (propagating modes) in the lead. The scattering matrix provides a linear relation between the incident and reflected waves $\ket{\text{out}} = S_E \ket{\text{in}}$. When considered as an analytic function of the complex energy $E$, the scattering matrix has a set of poles in the lower complex half-plane at energies $E_n - i\Gamma_n$ whose real and imaginary parts are random. The number of poles is the same as the number of eigenstates in the disordered sample when the measuring probe is detached. Coupling to the external lead makes these eigenstates metastable, thus shifting their energies in the lower complex half-plane of energy. Imaginary parts of the poles $\Gamma_n$ are directly related to the inverse decay time of these metastable states. After averaging over disorder realizations, we can describe the system by a distribution function of $\Gamma_n$. This distribution function will be the main object of our study.

So far, most analytic results for the distribution of scattering poles were obtained in the zero-dimensional limit (for relatively small systems at energies below $E_\text{Th}$) using random non-Hermitian matrices \cite{Haake1992, Fyodorov1999, Goetschy2011, Fyodorov2012}. In the unitary symmetry class, the problem was solved in great detail \cite{Fyodorov1997} in the framework of the nonlinear sigma model. Recently, some results for one-dimensional unitary systems (semi-infinite disordered wires) were also obtained \cite{Fyodorov2024}. A general integral formula for the zero-dimensional limit in the orthogonal class was derived in Ref.\ \cite{Sommers1999}. Experimental techniques to simulate complex quantum systems have been also developed in recent years. Quantum graphs \cite{Kottos2000, Kottos2003} based on microwave networks were used to model samples of all three classes: unitary \cite{Lawniczak2011, Bialous2016, Lawniczak2019, Lawniczak2020}, orthogonal \cite{Hul2004, Lawniczak2008, Lawniczak2012, Dietz2017}, and symplectic \cite{Rehemanjiang2016, Lawniczak2023, Lawniczak2024, Schmidt2024}. Chaotic quantum scattering was also simulated experimentally with microwave plane billiards \cite{Hlushchuk2000, Bialous2019}.

In this paper, we will use the nonlinear sigma model to develop a general and conceptually quite simple approach that equally well applies to all three Wigner-Dyson symmetry classes and to disordered systems of arbitrary geometry. Unlike previous studies, we will apply the sigma model to the whole setup including both the disordered part and the attached lead, see Fig.\ \ref{Fig:system}. Furthermore, we will introduce source terms in the sigma-model action, specifically designed to extract scattering matrix properties. It will allow us to derive a general integral expression for the distribution of scattering poles in all symmetry classes.

Based on this general integral formula, we will rederive and extend the classical result of Moldauer and Simonius \cite{Moldauer1967, Simonius1974} for the average decay rate. Our derivation is completely universal and shows that, in the classical limit, the average decay rate depends neither on the symmetry of the Hamiltonian nor on the geometry of the sample. We will then apply our general approach to study the zero-dimensional limit of the sigma model (disordered metallic grains or chaotic quantum scatterers). This will reproduce all the known analytic results for the unitary class \cite{Fyodorov1997}, significantly extend previous results in the orthogonal class \cite{Sommers1999} and provide a number of new results in the much less explored symplectic class. In all these cases, we will explicitly calculate arising integrals and derive concise analytic expressions for the distribution function of scattering poles.

The structure of the paper is the following. In Sec.\ \ref{Sec:formalism}, we formulate the problem and discuss some general properties of the scattering matrix. We then explain how to express the density of scattering poles in terms of the partition function of the sigma model with sources. In Sec.\ \ref{Sec:Moldauer-Simonius}, we apply our general formalism to solve the classical problem of the average decay rate \cite{Moldauer1967, Simonius1974}. We derive a universal result applicable to systems of any symmetry and geometry. In Sec.\ \ref{Sec:Fourier}, we compute the sigma-model partition function with the help of the Fourier transform on the sigma-model manifold \cite{Mirlin1994, KhalafDisser}. We apply this method to obtain a general integral formula for the distribution of scattering resonances. In Sec.\ \ref{Sec:0D}, we study the zero-dimensional limit of our general formula and derive a number of explicit distribution functions for small metallic grains of different symmetries. We then compare our results to numerical simulations of disordered systems in the framework of RMT in Sec.\ \ref{Sec:numerics}. In Sec.\ \ref{Sec:summary}, we summarize main results of the work and discuss possible extensions of our approach to solving other problems. Technical details related to the derivation of the sigma model with source terms and to the Fourier analysis on sigma-model manifolds are given in three Appendices.

\section{General formalism}
\label{Sec:formalism}

\subsection{Statement of the problem}

We consider a disordered sample attached to a semi-infinite one-dimensional clean ballistic wire as shown in Fig.\ \ref{Fig:system}. This ballistic wire hosts a fixed number $M$ of conducting channels in each direction. An electron wave function inside the wire is a $2M$-component vector that contains the amplitudes of left- and right-propagating modes. Reflection of electrons from the disordered sample can be described in most general terms by the energy-dependent unitary scattering matrix $S_E$ of size $M$. This matrix provides a linear relation between the amplitudes of incident and reflected plane waves in each channel.

Unitarity of the scattering matrix is the consequence of the particle conservation in the process of elastic scattering. Hence the determinant of $S_E$ is a complex number with the absolute value $1$. Another fundamental property of the scattering matrix can be deduced from its behavior as a function of the complex energy $E$. When analytically continued to complex energies, $S_E$ does not have singularities in the upper complex half-plane of $E$. This property is related to causality of scattering: a reflected wave packet leaves the system at a later time than the arrival of an incident wave packet. Using these two general properties of the scattering matrix, we can represent its determinant in the most general form as
\begin{equation}
 \det S_E
  = \prod_{n} \frac{E - E_n - i \Gamma_n}{E - E_n + i \Gamma_n}.
 \label{detS}
\end{equation}
This expression has manifestly the absolute value $1$ and is characterized by a set of poles in the lower complex half-plane at the energies $E_n - i \Gamma_n$. We can treat each pole as a metastable state inside the disordered system with the energy $E_n$ and the decay time $1/(2\Gamma_n)$. Finite decay times of the states appear due to the coupling to the external lead. In a closed system with the lead detached all $\Gamma_n$ would vanish. Our goal is to describe the distribution of these imaginary parts $\Gamma_n$ in a disordered metallic sample.

\subsection{Generating function}

As the first step towards solving the problem, we introduce the following generating function \cite{Fyodorov1997, Fyodorov2024}:
\begin{equation}
 F(E, \eta)
  = \left< \ln\det\left[ S^{-1}_{E - i\eta} S_{E + i\eta} \right] \right>.
 \label{F}
\end{equation}
Here the parameter $\eta$ is positive hence the two scattering matrices are taken at complex values of energy in the region of their respective analyticity. Angular brackets denote averaging with respect to disorder realizations. Using the decomposition (\ref{detS}), we can relate the generating function to the joint density of poles in the complex plane:
\begin{multline}
 \rho(E, \eta)
  = \frac{1}{4\pi} \left( \frac{\partial^2}{\partial E^2} + \frac{\partial^2}{\partial\eta^2} \right) F(E, \eta) \\
  = \left< \sum_n \delta(E - E_n) \delta(\eta - \Gamma_n) \right>.
 \label{rhoEF}
\end{multline}
In a metallic (or, in general, any semiclassical) sample, the density of electron states near the Fermi level is mostly a constant whose energy dependence can be completely disregarded. Neither disorder nor coupling to an external lead can significantly alter the value of this density. Hence the energy derivative in Eq.\ (\ref{rhoEF}) is very small and can be fully neglected. We simply replace the first delta-function factor in Eq.\ (\ref{rhoEF}) by the inverse level spacing $\Delta$ in the disordered sample. As a result, our generating function will depend only on $\eta$ and the density of poles becomes
\begin{equation}
 \rho(\eta)
  = \frac{\Delta}{4\pi} \frac{\partial^2}{\partial\eta^2} F(\eta)
  = \left< \sum_n \delta(\eta - \Gamma_n) \right>.
 \label{rhoF}
\end{equation}

\subsection{Source field}

Our next goal is to express the generating function (\ref{F}) in terms of the Green functions of the system. In order to do this, we will use a special form of the matrix Green function introduced by Nazarov in Ref.\ \cite{Nazarov1994}. This matrix Green function combines standard retarded and advanced Green functions of the system in a single object. In addition, it also contains a special source parameter $a$ known as the counting field. For our problem, the matrix Green function is defined as 
\begin{equation}
 \check G
  = \begin{pmatrix}
      E + i\eta - H & \hat v \sin(a/2) \delta(x - x_1) \\
      \hat v \sin(a/2) \delta(x - x_2) & E - i\eta - H
    \end{pmatrix}^{-1}.
 \label{checkG}
\end{equation}
Here $H$ is the full Hamiltonian of the system describing both the disordered part and the ballistic lead attached to it. The off-diagonal source terms are inserted at two points $x_{1,2}$ inside the lead and contain the velocity operator $\hat v$. The partition function (\ref{F}) can be directly written in terms of the determinant of $\check G$ as
\begin{equation}
 F
  = -\Bigl< \ln\det \check G(x_{1,2} = 0, a = \pi) \Bigr>.
 \label{FcheckG}
\end{equation}
This identity is derived in Appendix \ref{App:Nazarov}.

We see that the generating function $F$ is identical to the average free energy of the system in the fictitious external source field $a$. This allows us to apply standard methods of the theory of disordered metals to perform disorder averaging and hence to extract the average distribution function $\rho(\eta)$.

In order to average over disorder realizations, we should somehow get rid of the logarithm in Eq.\ (\ref{FcheckG}). Two standard approaches to resolving this issue are the replica trick and the supersymmetry. We will apply the latter approach and introduce two identical copies of the system: one with scattering bosons and one with fermions. Both the fermionic and bosonic copies of the system have identical realizations of disorder and behave in a completely similar fashion since the scattering is a single-particle elastic process. This means that we have two separate counting fields $a_\text{B,F}$ and the total partition function of the system is
\begin{equation}
 Z(a_\text{B,F})
  = \frac{\det \check G(a_\text{F})}{\det \check G(a_\text{B})}.
  \label{Z}
\end{equation}
When the two source fields are equal, $a_\text{B} = a_\text{F}$, we automatically have $Z = 1$, which is a manifestation of supersymmetry. The generating function $F$ can now be linearly related to the supersymmetric partition function $Z$ (no logarithm):
\begin{equation}
 F
  = \int_0^\pi da\, \frac{\partial \langle Z \rangle}{\partial a_\text{F}} \Bigr|_{a_\text{B,F} = a}.
 \label{FZ}
\end{equation}
This allows us to directly average the partition function over disorder.

\subsection{Nonlinear sigma model}

Average supersymmetric partition function of a disordered system can be calculated with the help of the nonlinear sigma model. This is a universal effective field theory written in terms of the supermatrix field $Q$ with the action \cite{Efetov1996}
\begin{equation}
 S[Q]
  = -\frac{\pi\nu}{4s} \int dr \operatorname{str} \left[  D (\nabla Q)^2 - 4\eta\Lambda Q \right].
 \label{sigma}
\end{equation}
Here $\nu$ is the density of states at the Fermi energy and $D$ is the diffusion constant. Matrix $Q$ obeys a nonlinear constraint $Q^2 = 1$ and can be represented as the rotated matrix $\Lambda$:
\begin{equation}
 Q
  = T^{-1} \Lambda T,
 \qquad
 \Lambda
  = \begin{pmatrix} 1 & 0 \\ 0 & -1 \end{pmatrix}_\text{RA}.
 \label{QLambda}
\end{equation}
In a disordered system without time-reversal symmetry (unitary class), $Q$ has the size $4 \times 4$ and acts in the direct product of the retarded-advanced space [same as for the matrix Green function $\check G$, see Eq.\ (\ref{checkG})] and Bose-Fermi superspace corresponding to the two copies of the disordered system. We will use shorthand notations RA and BF for these spaces, as we already did above in the expression for $\Lambda$. In the unitary class, the parameter $s$ in the action (\ref{sigma}) equals $1$.

Derivation of the sigma model is outlined in Appendix \ref{App:sigma}. The most nontrivial part of this derivation is the treatment of the source parameters $a_\text{B,F}$ in the lead \cite{Rejaei1996, Khalaf2016}. While the action in the disordered part of the system has its standard form (\ref{sigma}), the value of $Q$ inside the lead is fixed to a constant matrix
\begin{equation}
 Q_a
  = \begin{pmatrix} \cos \hat a & \sin \hat a \\ \sin \hat a & -\cos \hat a \end{pmatrix}_\text{RA},
 \quad
 \hat a
  = \begin{pmatrix} a_\text{B} & 0 \\ 0 & a_\text{F} \end{pmatrix}_\text{BF}.
 \label{Qa}
\end{equation}
Coupling between the value of $Q$ inside the disordered system and $Q_a$ in the lead is described by an additional boundary term in the action \cite{Efetov1996, Belzig2001}
\begin{gather}
 S_\Gamma(Q_0, Q_a)
  = \frac{1}{2s} \sum_n \operatorname{str} \ln \left[ g_n + \frac{1}{2} \bigl\{ Q_0, Q_a \bigr\} \right], \label{SGamma} \\
 g_n
  = \frac{2}{T_n} - 1. \label{gT}
\end{gather}
This boundary term contains a sum over conducting channels in the interface and parameters $T_n$ are transmission probabilities for individual channels. It turns out that a more convenient parameter for the channel transparency is $g_n$ as defined in Eq.\ (\ref{gT})

A disordered system with preserved time-reversal symmetry belongs to either orthogonal or symplectic symmetry class depending on whether the spin symmetry is preserved or broken. In both cases, matrix $Q$ has the size $8 \times 8$ and, in addition to RA and BF spaces, also acts in the particle-hole space (PH). We will denote Pauli matrices in this extra space by $\sigma_{x,y,z}$. For a system with time-reversal symmetry the parameter $s$ equals $2$ in the action (\ref{sigma}). The matrix $Q$ obeys an additional linear constraint depending on the type of the symmetry class:
\begin{equation}
 Q
  = C^T Q^T C,
 \qquad
 C
  = \begin{cases}
      \begin{pmatrix} \sigma_x & 0 \\ 0 & i\sigma_y \end{pmatrix}_\text{BF}, & \text{orthogonal}, \\
      \begin{pmatrix} i\sigma_y & 0 \\ 0 & \sigma_x \end{pmatrix}_\text{BF}, & \text{symplectic}.
    \end{cases}
\end{equation}

Full supersymmetric partition function of the system is given by a path integral with the full sigma-model action containing the bulk contribution (\ref{sigma}) and the boundary term (\ref{SGamma}). We will compute this integral in two steps. First, we integrate over all configurations $Q(\mathbf{r})$ inside the disordered part of the system with the fixed value $Q(\mathbf{r} = 0) = Q_0$ near the boundary with the external lead. This path integral defines the function
\begin{equation}
 \Psi(Q_0)
  = \int_{Q(\mathbf{r} = 0) = Q_0} DQ(\mathbf{r})\; e^{-S[Q(\mathbf{r})]}.
 \label{Psi}
\end{equation}
This function contains all the necessary information about electron dynamics in the disordered system and is sufficient to calculate any local observable at the point $\mathbf{r} = 0$. This is in particular true for the distribution of the poles of scattering matrix studied in the present paper. In the simplest case when the disordered system is a relatively small metallic grain, we can neglect spatial variations of $Q$. Setting the value of $Q$ at $\mathbf{r} = 0$ effectively fixes the matrix to the same value in the whole sample. Hence the function $\Psi$ takes the following simple form
\begin{equation}
 \Psi_\text{0D}(Q)
  = \exp\left[ -\frac{\pi \eta}{s \Delta} \operatorname{str} \left(\Lambda Q\right) \right].
 \label{Psi0D}
\end{equation}
Here $\Delta = (\nu V)^{-1}$ is the average level spacing ($V$ is the volume of the sample) and we have renamed $Q_0$ to just $Q$ for brevity. This form of the function $\Psi$ is valid for a disordered system of any geometry as long as $\eta$ is much lower than the Thouless energy $E_\text{Th}$. In the case of a semi-infinite disordered 1D wire, the function $\Psi_\text{1D}$ is known as the zero mode of the transfer matrix Hamiltonian \cite{Khalaf2017}. It is way more complicated but still can be found explicitly. In a more general context the function $\Psi$ is sometimes referred to as the order parameter function \cite{Fyodorov2024}.

Let us point out two important properties of the function $\Psi$. First, irrespective of the system geometry and of the value of $\eta$, the function is always normalized such that $\Psi(\Lambda) = 1$. This follows from the supersymmetry condition since $\Psi(\Lambda)$ is itself a full partition function of a disordered system coupled to an ideal metallic lead through a perfectly transparent interface with many conducting channels in the absence of any source parameters. Second, in the limit of large $\eta$ the potential term in the action (\ref{sigma}) is dominant and $\Psi$ decays fast as $Q$ deviates from $\Lambda$. This means that $\Psi$ effectively takes the form of a delta function on the sigma-model space
\begin{equation}
 \lim_{\eta \to \infty} \Psi(Q)
  = \delta(Q, \Lambda)
  = \begin{cases}
      1, & Q = \Lambda, \\
      0, & Q \neq \Lambda.
    \end{cases}
 \label{Psietainfty}
\end{equation}
At large but finite $\eta$, only values of $Q$ close to $\Lambda$ are relevant. One can expand the sigma-model action near this point and develop a systematic perturbation theory in terms of effective modes---diffusons and cooperons.

Once the function $\Psi(Q)$ is known, it remains to compute an ordinary integral over the values of $Q$ at the point $\mathbf{r} = 0$ in order to find the full average supersymmetric partition function (\ref{Z}). The integral contains two factors: the function $\Psi(Q)$ and the statistical weight associated with the boundary action (\ref{SGamma})
\begin{gather}
 \langle Z(a_\text{B,F}) \rangle
  = \int dQ\, \Psi(Q) \Gamma(Q, Q_a), \label{ZQ} \\
 \Gamma(Q, Q_a)
  = e^{-S_\Gamma(Q,Q_a)}. \label{Gamma}
\end{gather}

To summarize, the average distribution function $\rho(\eta)$ can be found in three steps: (i) calculate the convolution integral (\ref{ZQ}) for the partition function, (ii) apply Eq.\ (\ref{FZ}) to find the generating function $F(\eta)$, and (iii) extract the distribution function with the help of Eq.\ (\ref{rhoF}).

\section{Average decay rate}
\label{Sec:Moldauer-Simonius}

In the previous Section, we have developed a very general formalism to calculate the average distribution function of decay rates $\rho(\eta)$. Before dwelling into solving this problem in its full complexity, let us demonstrate how this general formalism yields the known classical result \cite{Moldauer1967, Simonius1974} for the average decay rate $\langle \eta \rangle$.

Using Eq.\ (\ref{rhoF}) and integrating by parts, we have
\begin{equation}
 \langle \eta \rangle
  = \int_0^\infty d\eta\, \eta\, \rho(\eta)
  = \frac{\Delta}{4\pi} \left[ \eta \frac{\partial F}{\partial\eta} - F \right]_{\eta = 0}^{\eta = \infty}.
\end{equation}
The first term in brackets vanishes at both ends of the integration interval. In the second term, we can apply Eqs.\ (\ref{FZ}) and (\ref{ZQ}) with the limiting forms of $\Psi$ at small and large $\eta$, Eqs.\ (\ref{Psi0D}) and (\ref{Psietainfty}). This yields
\begin{equation}
 \langle \eta \rangle
  = \frac{\Delta}{4\pi} \int_0^\pi\!\! da\, \frac{\partial}{\partial a_\text{F}} \int dQ \Bigl[ 1 - \delta(Q, \Lambda) \Bigr] \Gamma(Q, Q_a) \Bigr|_{a_\text{B,F} = a}.
\end{equation}
We interchange the order of integrals first applying the derivative in $a_\text{F}$ and the integral over $a$ directly to $\Gamma$ and only then integrating over $Q$. Contribution of the first term in brackets (unity) vanishes. This follows from the symmetry of the sigma-model manifold. Indeed, the result of integration of $\Gamma(Q,Q_a)$ over all values of $Q$ is independent of $Q_a$ since all points of the manifold are equivalent. Hence the derivative in $a_\text{F}$ cancels this contribution. Only the second term in brackets (the delta function) contributes to the final result. It yields
\begin{equation}
 \langle \eta \rangle
  = -\frac{\Delta}{4\pi} \int_0^\pi da\, \frac{\partial}{\partial a_\text{F}} \Gamma(\Lambda, Q_a) \Bigr|_{a_\text{B,F} = a}.
 \label{etaGamma}
\end{equation}
Using the explicit form of $\Gamma$ from Eqs.\ (\ref{SGamma}) and (\ref{Gamma}) and the source matrix $Q_a$ from Eq.\ (\ref{Qa}), we have
\begin{equation}
 \Gamma(\Lambda, Q_a)
  = \prod_{n = 1}^M \frac{g_n + \cos a_\text{F}}{g_n + \cos a_\text{B}}.
 \label{GammaQa}
\end{equation}
Then Eq.\ (\ref{etaGamma}) provides the result
\begin{equation}
 \langle \eta \rangle
  = \frac{\Delta}{4\pi} \sum_n \ln \frac{g_n + 1}{g_n - 1}
  = -\frac{\Delta}{4\pi} \sum_n \ln (1 - T_n).
 \label{aveta}
\end{equation}

This expression for the average decay rate in terms of transparencies $T_n$ of individual channels is known as the Moldauer-Simonius relation \cite{Moldauer1967, Simonius1974}. It was first derived for a system with time-reversal symmetry by averaging over the spectrum. The same result with disorder average in the system with broken time-reversal symmetry was found by Sommers and Fyodorov \cite{Fyodorov1997}. We have just shown that the expression (\ref{aveta}) is fully universal and depends neither on symmetry nor on geometry of the disordered system. This signifies the semiclassical nature of the average decay rate.

We also note that the average decay rate diverges logarithmically when channel transparencies $T_n$ approach unity. From this property we can infer the large $\eta$ tail of the distribution function $\rho(\eta)$ in the case of $M$ perfectly transparent channels:
\begin{equation}
 \rho(\eta \gg \Delta)
  \approx \frac{M \Delta}{4\pi\eta^2}.
 \label{largeetatail}
\end{equation}
This tail is also semiclassical and universal. We will demonstrate below that exact distribution functions in small metallic grains of any symmetry do possess this large $\eta$ tail.

\section{Fourier analysis on the symmetric superspace}
\label{Sec:Fourier}

Previously, we have reduced the problem of finding the distribution function (\ref{rhoF}) to the calculation of the convolution integral (\ref{ZQ}) on the sigma-model manifold. In the present Section, we will use geometric properties of the manifold to significantly simplify this integral.

The supermatrix $Q$ can be represented as a rotated matrix $\Lambda$, see Eq.\ (\ref{QLambda}). The rotation matrix $T$ belongs to a certain supergroup, $T \in G$, whose exact structure depends on the particular symmetry class of the problem, see Table \ref{Tab:classes}. Note that the bosonic/fermionic parts of the supergroup $G$ are always noncompact/compact, respectively.

\begin{table*}
\begin{center}
\begin{tblr}{Q[c,m]Q[c,m]Q[c,m]Q[c,m]Q[c,m]Q[c,m]Q[c,m]Q[c,m]}
\hline\hline
Class & $s$ & $w$ & $G$ & $K$ & $\hat\theta$ & $\theta$ domain & $\mathbf{q}$ \\
\hline
unitary
 & $1$
 & $1$
 & $\mathrm{U}(1, 1 \vert 2)$
 & $\mathrm{U}(1 \vert 1) \times \mathrm{U}(1 \vert 1)$
 & $\begin{pmatrix} \theta_\text{B} & 0 \\ 0 & i \theta_\text{F} \end{pmatrix}_\text{BF}$
 & {$\theta_\text{B} > 0$ \\ $0 < \theta_\text{F} < \pi$}
 & {$\{q,\, l\}$ \\ $l = 0,1,2,\ldots$ \\ $q > 0$}
\\
orthogonal
 & $2$
 & $2$
 & $\mathrm{OSp}(2, 2 \vert 4)$
 & $\mathrm{OSp}(2 \vert 2) \times \mathrm{OSp}(2 \vert 2)$
 & $\begin{pmatrix} \theta_\text{B} + \sigma_x \theta_\text{B2} & 0 \\ 0 & i \theta_\text{F} \end{pmatrix}_\text{BF}$
 & {$\theta_\text{B,B2} > 0$ \\ $0 < \theta_\text{F} < \pi$}
 & {$\{q,\, q_2,\, l\}$ \\ $l = 0,1,2\ldots$ \\ $q, q_2 > 0$}
\\
symplectic
 & $2$
 & $1$
 & $\mathrm{SpO}(2, 2 \vert 4)$
 & $\mathrm{SpO}(2 \vert 2) \times \mathrm{SpO}(2 \vert 2)$
 & $\begin{pmatrix} \theta_\text{B} & 0 \\ 0 & i \theta_\text{F} + i \sigma_x \theta_\text{F2} \end{pmatrix}_\text{BF}$
 & {$\theta_\text{B} > 0$ \\ $0 < \theta_\text{F2} < \theta_\text{F} < \pi$}
 & {$\{q,\, l,\, l_2\}$ \\ $l, l_2 = 0,1,2,\ldots$ \\ $l + l_2$ even, $q > 0$}
\\
\hline\hline
\end{tblr}
\end{center}
\caption{
Parameters of the sigma-model manifolds in the three Wigner-Dyson symmetry classes. The normalization constant $s$ is related to the size of the $Q$ matrix and appears in Eqs.\ (\ref{sigma}) and (\ref{SGamma}). The parameter $w$ arises in the analytic continuation, Eqs.\ (\ref{tGamma}) and (\ref{tGammalambda}). Supergroup $G$ and its subgroup $K$ define the sigma-model manifold: $Q \in G/K$. Cartan angles $\theta$ are introduced in the parametrization of $Q$ in Eqs.\ (\ref{Cartan}) and (\ref{Theta}). The momentum parameters in the vector $\mathbf{q}$ enumerate irreducible representations of $G$ and hence label zonal spherical functions $L_\mathbf{q}$ [eigenfunctions of the radial Laplace-Beltrami operator (\ref{Delta})] on the sigma-model manifold. 
}
\label{Tab:classes}
\end{table*}

Two different matrices $T_1$ and $T_2$ define the same matrix $Q$ if they are related by a left rotation $T_2 = K T_1$ with the matrix $K$ from the same group $G$ that in addition commutes with $\Lambda$. Such matrices form a compact subgroup $K \subset G$ also listed in Table \ref{Tab:classes}. Distinct matrices $Q$ correspond to right cosets of $K$ in $G$. This coset space has the structure of a symmetric superspace and plays the role of the sigma-model manifold: $Q \in G/K$. One can think of this manifold as a generalization of an ordinary sphere to a multidimensional superspace. Much like a sphere is generated by usual $\mathrm{SO}(3)$ rotations of a unit vector, a symmetric superspace is generated by the rotations $T$ of a ``unit'' matrix $\Lambda$ according to Eq.\ (\ref{QLambda}).

It is very convenient to introduce special coordinates on the sigma-model manifold \cite{Efetov1996} by means of the Cartan decomposition of the matrix $T$. We can represent $Q$ in the form
\begin{equation}
 Q
  = U^{-1} \Lambda e^{\Theta} U,
 \qquad
 U
  \in K.
 \label{Cartan}
\end{equation}
Using a rotation with the matrix $U$ from the group $K$, we bring $Q$ to an ``almost diagonal'' form. Namely, the only remaining parameters after $U$ rotation are contained in the matrix $\Theta$ that belongs to the algebra of $G$, is abelian, and anticommutes with $\Lambda$. In all three symmetry classes we choose the same RA structure of $\Theta$:
\begin{equation}
 \Theta
  = \begin{pmatrix} 0 & -i \hat\theta \\ i\hat\theta & 0 \end{pmatrix}_\text{RA}.
 \label{Theta}
\end{equation}
The block $\hat\theta$ acts in the remaining BF and PH (in the orthogonal and symplectic classes) spaces and contains two or three parameters depending on the symmetry class, see Table \ref{Tab:classes}. In all three classes, two of these parameters, $\theta_\text{B,F}$, have the same matrix structure as the source fields $a_\text{B,F}$ in Eq.\ (\ref{Qa}). The only difference is that the bosonic angle $\theta_\text{B}$, unlike the source parameter $a_\text{B}$, is noncompact. We will routinely assume the matrix $Q_a$ inside the metallic lead to be from the same class as the matrix $Q$ in the disordered part of the system. Identification of the two matrices implies analytic continuation in the bosonic variable:
\begin{equation}
 \theta_\text{B} \leftrightarrow i a_\text{B},
 \qquad
 \theta_\text{F} \leftrightarrow a_\text{F},
 \qquad
 \theta_\text{B2/F2} \leftrightarrow 0.
 \label{atheta}
\end{equation}

Cartan parametrization (\ref{Cartan}) of the matrix $Q$ is a generalization of the usual spherical coordinates on an ordinary sphere. Rotations by the matrix $U$ leave the point $\Lambda$ of the manifold unchanged exactly like rotations by the azimuthal angle do not shift the position of the north pole on a sphere. Matrix $U$ is thus analogous to the longitude. Parameters of the matrix $\hat\theta$ measure the ``distance'' between $Q$ and $\Lambda$ similar to how the polar angle measures the distance from the north pole (latitude) on a usual sphere.

The function $\Psi(Q)$ defined by the path integral (\ref{Psi}) with the action (\ref{sigma}) is invariant with respect to any rotations of $Q$ by the group $K$: 
\begin{equation}
 \Psi(K^{-1} Q K)
  = \Psi(Q)
  = \Psi(\theta).
\end{equation}
Here we simply write the argument $\theta$ to denote the full set of two or three parameters of the matrix $\hat\theta$ listed in Table \ref{Tab:classes}. Similarly, the function $\Gamma$ defined in Eq.\ (\ref{Gamma}) is invariant with respect to a simultaneous rotation of its both arguments by the same matrix from the group $G$:
\begin{gather}
 \Gamma(G^{-1} Q G, G^{-1} Q_a G)
  = \Gamma(Q, Q_a)
  = \Gamma(\tilde\theta), \\
 \tilde Q
  = T Q_a T^{-1}.
\end{gather}
In the last expression, we have defined the ``relative'' matrix $\tilde Q$ that is the result of the inverse rotation $T^{-1}$ from Eq.\ (\ref{QLambda}) applied to the matrix $Q_a$. The function $\Gamma$ depends only on the Cartan angles $\tilde\theta$ of this ``relative'' matrix $\tilde Q$.

\begin{figure}
\includegraphics[width=0.25\textwidth]{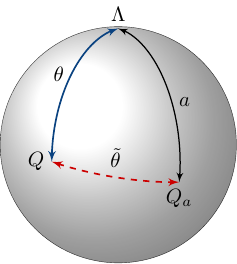}
\caption{An illustration of the sigma-model manifold by an ordinary two-dimensional sphere. The matrix $\Lambda$ corresponds to the north pole. Matrices $Q$ and $Q_a$ are represented as points on the sphere with the polar angles $\theta$ and $a$, respectively. Angular distance between $Q$ and $Q_a$ is $\tilde\theta$.}
\label{Fig:sphere}
\end{figure}

We can illustrate geometric meaning of the notations introduced above by an examle of a usual sphere shown in Fig.\ \ref{Fig:sphere}. The point $\Lambda$ has the meaning of the north pole on the sphere. The function $\Psi$ depends only on the angular distance from the north pole to the point $Q$ on the sphere (latitude of $Q$). The function $\Gamma$ depends only on the angular distance $\tilde\theta$ between points $Q$ and $Q_a$. Convolution of the two functions (\ref{ZQ}) is the integral with respect to $Q$ of their product over the whole sphere. The result of integration depends only on the angular distance $a$ between $\Lambda$ and $Q_a$.

\subsection{Spherical functions}

We have so far established that the integral (\ref{ZQ}) is a convolution of two functions on the sigma-model manifold. Both functions depend only on the ``distance'' between corresponding points. Such convolution integrals are most easily computed with the help of the Fourier transform. In the case of a symmetric space, the basis for the Fourier expansion is a full set of spherical functions---eigenfunctions of the Laplace-Beltrami operator on the corresponding manifold. In the case of a usual sphere, they are the standard spherical functions while the Laplace-Beltrami operator is the square of the angular momentum. Moreover, since we are interested in the convolution of two functions that depend only on the angular distance, their Fourier expansion involves only the so called zonal spherical functions. On an ordinary sphere these are the functions with zero projection of the angular momentum on the vertical axis. Such functions depend only on the polar but not on the azimuthal angle.

Zonal spherical functions on the sigma-model manifold are known explicitly for all three Wigner-Dyson symmetry classes \cite{Mirlin1994, KhalafDisser}. They are the eigenfunctions of the radial part (acting only on $\theta$ variables) of the Laplace-Beltrami operator
\begin{equation}
 \Delta
  = \frac{1}{J(\theta)} \sum_i \frac{\partial}{\partial\theta_i} J(\theta) \frac{\partial}{\partial\theta_i}.
 \label{Delta}
\end{equation}
Summation index $i$ here takes two or three values depending on the number of $\theta$ variables in the corresponding symmetry class. Jacobian function $J(\theta)$ for all classes is given in Appendix \ref{App:eigenfunctions}. Spherical functions are labeled by a vector index $\mathbf{q}$ that we will simply refer to as momentum. It has both discrete and continuous components for the compact and noncompact variables, respectively. In every class there is at least one continuous momentum component $q > 0$ in the bosonic sector and one discrete component $l = 0,1,2,\ldots$ in the fermionic sector. Full structure of $\mathbf{q}$ is shown in Table \ref{Tab:classes}. There is also a special eigenfunction that equals the constant $1$ on the whole manifold and has the eigenvalue $0$. We will not assign any value of $\mathbf{q}$ to this specific function.

We will denote eigenfunctions of the Laplace-Beltrami operator (\ref{Delta}) by $L_\mathbf{q}(\theta)$; they are given explicitly in Appendix \ref{App:eigenfunctions}. Spherical functions are normalized such that the following two identities hold:
\begin{subequations}
\label{normalization}
\begin{gather}
 \int d\theta\, J(\theta)\, L_\mathbf{q}(\theta)\, L_\mathbf{q'}(\theta)
  = \delta_{\mathbf{q}, \mathbf{q'}}, \\
 1 + \sum_\mathbf{q} L_\mathbf{q}(\theta)
  = \delta(Q, \Lambda)
  = \begin{cases} 1, & \theta = 0, \\ 0, & \theta \neq 0. \end{cases} \label{normalization_delta}
\end{gather}
\end{subequations}
The summation symbol here implies a sum/integral over discrete/continuous components of $\mathbf{q}$ with an appropriate measure also given in Appendix \ref{App:eigenfunctions}. Kronecker symbol in the first identity also implies the same measure. The second identity represents a Fourier expansion of the delta function $\delta(Q, \Lambda)$ on the sigma-model manifold. This delta function equals $1$ when $Q = \Lambda$ and $0$ otherwise. We normalize our spherical functions such that all the Fourier components of the delta function are $1$.

Once the proper basis of the zonal spherical functions on the sigma-model manifold is established, we can represent both $\Psi$ and $\Gamma$ by their Fourier series:
\begin{subequations}
\label{Fourier}
\begin{gather}
 \Psi(\theta)
  = \psi_0 + \sum_\mathbf{q} \psi_\mathbf{q} L_\mathbf{q}(\theta), \\
 \Gamma(\tilde\theta)
  = \gamma_0 + \sum_\mathbf{q} \gamma_\mathbf{q} L_\mathbf{q}(\tilde\theta). \label{GammaFourier}
\end{gather}
\end{subequations}
Then the average partition function given by the convolution integral (\ref{ZQ}) can be written as
\begin{equation}
 \langle Z(a) \rangle
  = \psi_0\gamma_0 + \sum_\mathbf{q} \psi_\mathbf{q} \gamma_\mathbf{q} L_\mathbf{q}(a).
 \label{ZFourier}
\end{equation}

We can prove this identity by the following simple argument. First, since the function $\langle Z \rangle$ depends only on the Cartan variables $a$ of the matrix $Q_a$, it can be expanded in the zonal spherical functions of $a$ with some coefficients. Each zonal spherical function belongs to a distinct irreducible representation of the group $G$ acting on $Q$ and the index $\mathbf{q}$ labels these representations. Hence a convolution of two such spherical functions is nonzero only if they belong to the same representation and have the same index $\mathbf{q}$. This proves that each term in the sum in Eq.\ (\ref{ZFourier}) contains the product of Fourier amplitudes $\psi_\mathbf{q} \gamma_\mathbf{q}$ with the same index. Finally, the fact that there are no other $\mathbf{q}$-dependent factors in the Fourier expansion (\ref{ZFourier}) follows from the normalization condition (\ref{normalization_delta}). Indeed, if we substitute $\delta(Q,\Lambda)$ in the place of $\Psi(Q)$ in Eq.\ (\ref{ZQ}), it will yield $\langle Z(a) \rangle = \Gamma(a)$. The same result will follow from the Fourier series (\ref{ZFourier}) if we insert the Fourier amplitudes $\psi_0 = \psi_\mathbf{q} = 1$ of the delta function there. We have thus established that the average partition function (\ref{ZQ}) is indeed given by the Fourier series (\ref{ZFourier}).

\subsection{Effect of the source field}

The Fourier expansion (\ref{ZFourier}) of the partition function can be directly used in Eq.\ (\ref{FZ}) to compute $F(\eta)$. We apply the linear operator from Eq.\ (\ref{FZ}) to each individual spherical function and obtain a $\mathbf{q}$-dependent number
\begin{equation}
 \phi_\mathbf{q}
  = \int_0^\pi da\, \frac{\partial L_\mathbf{q}(\theta)}{\partial \theta_\text{F}} \Bigr|_{\substack{\theta_\text{B} = i\theta_\text{F} = i a\\ \theta_\text{B2/F2} = 0}}.
\end{equation}
Let us remind that the arguments of the function $L_\mathbf{q}(a)$ in Eq.\ (\ref{ZFourier}) correspond to the Cartan variables $\theta$ according to the convention (\ref{atheta}).

Using explicit formulas for spherical functions from Appendix \ref{App:eigenfunctions}, we can find $\phi_\mathbf{q}$ for all three symmetry classes
\begin{multline}
 \phi_\mathbf{q}
  = \frac{(-1)^{l+1}}{\pi} \cosh(\pi q) \\
    \times \begin{dcases}
     1, & \text{unitary}, \\
     2, & \text{orthogonal}, \\
     1 - \frac{(l - l_2)(l + l_2 + 1)}{q^2 + 1/4}, & \text{symplectic}.
    \end{dcases}
 \label{phiq}
\end{multline}
With this result, we represent the generating function as a sum over Fourier harmonics:
\begin{equation}
 F(\eta)
  = \sum_\mathbf{q} \psi_\mathbf{q} \gamma_\mathbf{q} \phi_\mathbf{q}.
 \label{Fq}
\end{equation}

This form of the generating function is not much easier for practical calculations than the original expression for $\langle Z \rangle$ with the convolution integral (\ref{ZQ}) and subsequent application of Eq.\ (\ref{FZ}). It can be used only once the Fourier expansions (\ref{Fourier}) are explicitly found, which itself is quite a hard computational task. Below, we will present a much more efficient formula for the generating function.

\subsection{Analytic continuation}

The polar angle $\theta$ on a usual sphere is defined in the interval $[0, \pi]$. At the same time, each spherical function is given by a Legendre polynomial $P_l(\cos\theta)$ \cite[\href{https://dlmf.nist.gov/14.3.E6}{(14.3.6)}]{DLMF} and hence can be extended to all values of $\theta$ with the period $2\pi$. Functions with even/odd values of $l$ are even/odd under the shift by half a period $\theta \mapsto \theta + \pi$. This is a well-known property of the spherical functions related to their parity under spatial inversion. The shift of $\theta$ by $\pi$ indeed maps each point on the sphere on to the opposite point.

A similar property for the eigenfunctions on the noncompact analogue of a sphere---hyperboloid---is a bit trickier. Polar angle $\theta$ on a hyperboloid is defined on the half-axis $\theta > 0$. Spherical functions (also known as conical functions) are labeled by a continuous variable $q$ and have the form $P_{-1/2 + i q}(\cosh\theta)$. They are periodic in the complex plane of $\theta$ with the period $2\pi i$. These functions have a branch cut discontinuity inside each period running along the line with $\operatorname{Im} \theta$ being an odd multiple of $\pi$. If we perform analytic continuation in the upper/lower complex half-plane $\theta \mapsto \theta \pm i\pi$, the spherical function will take two different values corresponding to the two sides of the branch cut. The difference between these two values (jump of the function across the branch cut) equals to the original value of the function up to multiplication by $2i\cosh(\pi q)$. This identity follows from the properties of Legendre functions and is a counterpart for the parity of spherical functions on a usual compact sphere.

We can generalize these parity identities to the spherical functions on the sigma-model manifold. Inspecting explicit expressions for the spherical functions from Appendix \ref{App:eigenfunctions} and using the properties of Legendre functions, we find the following relation:
\begin{equation}
 \operatorname{Im} L_\mathbf{q}(\theta_\text{B} + i \pi, \theta_\text{F} + \pi)
  = (-1)^l \cosh(\pi q) L_\mathbf{q}(\theta).
 \label{parity}
\end{equation}
This simple identity universally applies to all three Wigner-Dyson symmetry classes.

We observe a striking similarity between the coefficient in the right-hand side of Eq.\ (\ref{parity}) and the value of $\phi_\mathbf{q}$ from Eq.\ (\ref{phiq}). In the unitary and orthogonal classes, the two quantities are equal up to a constant factor. It turns out that in the symplectic class both functions $\Psi$ and $\Gamma$ are symmetric under the interchange $\theta_\text{F} \leftrightarrow \theta_\text{F2}$ hence their Fourier components are also symmetric under $l \leftrightarrow l_2$. In view of this property, we can also use the symmetrized version of $\phi_\mathbf{q}$ in Eq.\ (\ref{Fq}) for the symplectic class. Such symmetrized $\phi_\mathbf{q}$ is exactly the same as in the unitary class and also coincides up to a constant with the factor in Eq.\ (\ref{parity}).

We define a new function $\tilde\Gamma(\theta)$ as the analytic continuation of the type of Eq.\ (\ref{parity}) applied to the function $\Gamma(\theta)$. Using the parity property of the spherical harmonics, the Fourier series (\ref{GammaFourier}) for this new function is
\begin{equation}
 \tilde\Gamma(\theta)
  = -\frac{w}{\pi} \operatorname{Im} \Gamma(\theta_\text{B} + i \pi, \theta_\text{F} + \pi)
  = \sum_\mathbf{q} \gamma_\mathbf{q} \phi_\mathbf{q} L_\mathbf{q}(\theta).
 \label{tGamma}
\end{equation}
Here the factor $w$ equals $2$ in the orthogonal class and $1$ in the unitary and symplectic classes, see Table \ref{Tab:classes}. In the latter case, we also assumed symmetrized versions of $\phi_\mathbf{q}$ and $L_\mathbf{q}(\theta)$ with respect to $l \leftrightarrow l_2$ in the right-hand side.

After this analytic continuation, the Fourier sum in Eq.\ (\ref{Fq}) represents an overlap of two functions on the sigma-model manifold. We can perform the inverse Fourier transform and rewrite this overlap as an integral in the original variables:
\begin{equation}
 F(\eta)
  = \int d\theta J(\theta) \Psi(\theta) \tilde\Gamma(\theta).
 \label{FPsiGamma}
\end{equation}
This integral is dramatically simpler than the original convolution integral (\ref{ZQ}). First, it does not involve any source parameters and provides the generating function $F(\eta)$ directly. Second, it implies integration over the $\theta$ variables only rather than over the whole sigma-model manifold. In the orthogonal and symplectic classes, we have thus reduced an integral over eight commuting and eight Grassmann variables to just a three-dimensional integral over real numbers.

It remains to take a derivative in $\eta$ in Eq.\ (\ref{rhoF}) to find the distribution function of scattering poles. We can do this directly in Eq.\ (\ref{FPsiGamma}) and get the final expression
\begin{equation}
 \rho(\eta)
  = \frac{\Delta}{4 \pi} \int d\theta J(\theta) \frac{\partial^2 \Psi(\theta)}{\partial\eta^2} \tilde\Gamma(\theta).
 \label{rhoPsiGamma}
\end{equation}

This integral representation of the distribution function is the central result of our work. Let us stress that it universally applies to all three symmetry classes and allows for any geometry of the disordered system as long as the function $\Psi$ is known. We have also incorporated the most general boundary conditions for the coupling between the disordered system and the external measuring probe encoded in the function $\Gamma$, see Eq.\ (\ref{Gamma}). The integral (\ref{rhoPsiGamma}) was derived with the help of the Fourier transform on the sigma-model manifold, but the final form of the result does not require any knowledge of spherical functions and does not imply any complicated Fourier integration.

The distribution of scattering poles in small metallic grains of unitary symmetry was first derived in Ref.\ \cite{Fyodorov1997}. The technically much harder case of orthogonal symmetry class was studied in Ref.\ \cite{Sommers1999}. Later, a formula similar to Eq.\ (\ref{rhoPsiGamma}) for the unitary symmetry class was obtained in Ref.\ \cite{Fyodorov2024}, which allowed to find the distribution of scattering resonances in semi-infinite disordered wires. To the best of our knowledge, neither the symplectic symmetry class nor a general and universal formula of the form (\ref{rhoPsiGamma}) applicable to all classes were ever discussed before. Below, we will use Eq.\ (\ref{rhoPsiGamma}) to reproduce all the known results for small metallic grains in the unitary class, significantly extend previously known results for the orthogonal class and present new results for the symplectic class.

\section{Distribution of scattering poles in disordered metallic grains}
\label{Sec:0D}

In this Section, we will apply the general integral representation (\ref{rhoPsiGamma}) to study the distribution of scattering poles in small metallic grains. As we have discussed earlier, these results are also directly applicable to chaotic systems thanks to the Bohigas-Giannoni-Schmidt conjecture \cite{Bohigas1984}.

In a small grain, gradients of $Q$ are negligible hence the function $\Psi$ is given by Eq.\ (\ref{Psi0D}). We will also measure $\eta$ in units of mean level spacing in the grain and introduce a dimensionless energy variable
\begin{equation}
 y
  = \frac{2\pi\eta}{\Delta}.
 \label{y}
\end{equation}
To make further equations more compact, we will also use trigonometric variables \cite{Efetov1996} instead of the original $\theta$ angles:
\begin{gather}
 \lambda_\text{B,B2}
  = \cosh\theta_\text{B,B2},
 \qquad
 \lambda_\text{F,F2}
  = \cos\theta_\text{F,F2}, \label{lambda} \\
 \lambda_\text{B,B2} > 1, \qquad -1 < \lambda_\text{F} < \lambda_\text{F2} < 1. 
\end{gather}

In terms of these new variables, basic equations (\ref{tGamma}) and (\ref{rhoPsiGamma}) take the form
\begin{gather}
 \tilde\Gamma(\lambda)
  = -\frac{w}{\pi} \operatorname{Im} \Gamma(-\lambda_\text{B} + i0, -\lambda_\text{F}), \label{tGammalambda} \\
 \rho(y)
  = \frac{1}{2} \int d\lambda\, J(\lambda) \frac{\partial^2 \Psi(\lambda)}{\partial y^2} \tilde\Gamma(\lambda). \label{rhoPsiGammalambda}
\end{gather}
To reduce unnecessary technical details, we will focus on the case of $M$ identical channels in the measuring lead all having the same transparency parameter $g = 2/T - 1$.

\subsection{Unitary class}

In the unitary symmetry class, the matrix $Q$ has one bosonic and one fermionic Cartan variable corresponding to $\lambda_\text{B,F}$ according to Eq.\ (\ref{lambda}). The function $\Gamma$ that describes the coupling to the lead is given by Eqs.\ (\ref{SGamma}) and (\ref{Gamma}). In the most general case it is quite similar to Eq.\ (\ref{GammaQa})
\begin{equation}
 \Gamma(\lambda)
  = \prod_{n = 1}^M \frac{g_n + \lambda_\text{F}}{g_n + \lambda_\text{B}}.
 \label{GammaA}
\end{equation}
After analytic continuation and taking the imaginary part according to Eq.\ (\ref{tGammalambda}), we obtain a sum of delta functions for the bosonic variable, $\delta(\lambda_\text{B} - g_n)$, with the fixed value in each individual channel. These delta functions arise due to an infinitesimal imaginary part $i0$ entering the denominator of the fraction in Eq.\ (\ref{GammaA}) after analytic continuation. The delta functions allow us to reduce Eq.\ (\ref{rhoPsiGammalambda}) to an integral over $\lambda_\text{F}$ only. This most general case for the distribution function in a small metallic grain of the unitary symmetry class was studied in detail in Ref.\ \cite{Fyodorov1997}.

\subsubsection{General result for identical channels}

We will restrict our analysis here to a simpler case of $M$ identical conducting channels with the same transparency $g$. To consider all possible values of $M$ at once, we define a single function $\Gamma$ with an auxiliary parameter $\alpha$ as
\begin{equation}
 \Gamma(\alpha, \lambda)
  = \sum_{M = 1}^\infty (-\alpha)^{M-1} \Gamma_M(\lambda)
  = \frac{g + \lambda_\text{F}}{g + \lambda_\text{B} + \alpha(g + \lambda_\text{F})}.
 \label{Gammaalpha}
\end{equation}
Later, the distribution for any given value of $M$ can be extracted by expanding in powers of small $\alpha$ to the appropriate order.

Performing analytic continuation and taking imaginary part of $\Gamma$ according to Eq.\ (\ref{tGammalambda}) yields
\begin{equation}
 \tilde\Gamma(\alpha, \lambda)
  = (g - \lambda_\text{F}) \delta\bigl[ g - \lambda_\text{B} + \alpha(g - \lambda_\text{F}) \bigr].
\end{equation}
Here we have only a single delta function for any number of channels but its position is shifted away from the point $\lambda_\text{B} = g$. The delta function allows us to take the integral over $\lambda_\text{B}$ in Eq.\ (\ref{rhoPsiGammalambda}). The distribution function is then given by the following simple integral over the compact fermionic variable only:
\begin{equation}
 \rho(\alpha, y)
  = \frac{1}{2(1 + \alpha)^2} \int_{-1}^1 \frac{d\lambda_\text{F}}{g - \lambda_\text{F}} \frac{\partial^2 \Psi}{\partial y^2} \Bigr|_{\lambda_\text{B} = g + \alpha(g - \lambda_\text{F})}.
 \label{rhoA}
\end{equation}
This result in its general form with an arbitrary function $\Psi$ was first derived in Ref.\ \cite{Fyodorov2024} where it was used to study the distribution function in a semi-infinite disordered wire.

\begin{figure}
  \centerline{\includegraphics[width=\columnwidth]{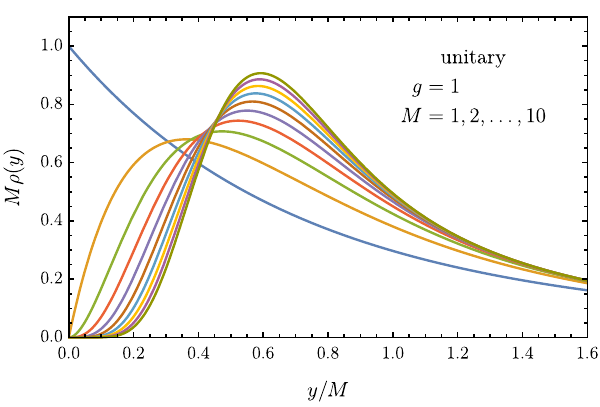}}
  \centerline{\includegraphics[width=\columnwidth]{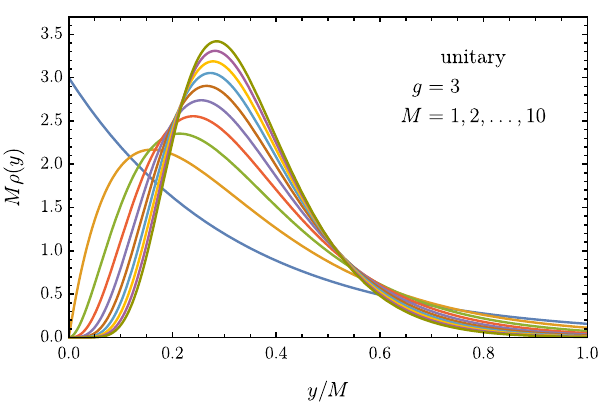}}
  \caption{The density of scattering poles $\rho(y)$ in a small metallic grain of the unitary class, Eq.\ (\ref{rhounitM}), with the attached measuring probe hosting $M = 1, \ldots 10$ conducting channels. Upper panel: perfect coupling with $g = 1$. Lower panel: each channel is coupled with $g = 3$ that corresponds to the transparency $T = 0.5$. We plot the distribution for the scaled variable $y/M$ to facilitate visual comparison of different curves.}
  \label{Fig:resA}
\end{figure}

Let us outline the results for the distribution function in small metallic grains of the unitary symmetry class. The function $\Psi$ in this case is given by Eq.\ (\ref{Psi0D}):
\begin{equation}
 \Psi
  = e^{-y(\lambda_\text{B} - \lambda_\text{F})}.
\end{equation}
With such a simple $\Psi$, the integral in Eq.\ (\ref{rhoA}) takes the form
\begin{equation}
 \rho(\alpha, y)
  = \frac{1}{2} \int_{-1}^1 d\lambda_\text{F}\, (g - \lambda_\text{F})\, e^{-y(1 + \alpha)(g - \lambda_\text{F})}.
 \label{rhounitint}
\end{equation}
It is now straightforward to expand the integrand in powers of small $\alpha$ and extract the distribution function for any number of channels $M$
\begin{multline}
 \rho_M(y)
  = \frac{y^{M-1}}{2(M - 1)!} \int_{-1}^1 d\lambda_\text{F}\, (g - \lambda_\text{F})^M\, e^{-y(g - \lambda_\text{F})} \\
  = \frac{\gamma[M + 1, (g + 1)y] - \gamma[M + 1, (g - 1)y]}{2y^2 (M - 1)!}.
 \label{rhounitM}
\end{multline}
This general result is written in terms of the lower incomplete gamma function \cite[\href{https://dlmf.nist.gov/8.2.E1}{(8.2.1)}]{DLMF}. It was first presented in Ref.\ \cite{Fyodorov1997} in a slightly different form. For any given value of $M$ it can be also expressed using elementary functions only. We show the distribution function (\ref{rhounitM}) for the first few values of $M$ in Fig.\ \ref{Fig:resA}.

\subsubsection{Asymptotics of large and small \texorpdfstring{$y$}{y}}

Let us assume for simplicity the limit of perfectly transparent channels $g = 1$. In this case, the distribution function has the following asymptotic behavior at small and large values of $y$:
\begin{equation}
 \rho_M(y)
  \approx \begin{dcases}
      \frac{2^M M y^{M - 1}}{(M + 1)!}, & y \ll 1, \\
      \frac{M}{2y^2}, & y \gg 1.
    \end{dcases}
 \label{rhounitasymp}
\end{equation}
The large $y$ tail corresponds to the classical result (\ref{largeetatail}) discussed above. Corrections to this asymtotic form are exponentially small. At low values of $y$, we have a strongly quantum limit of the distribution. The probability to find a quantum state that is very weakly coupled to the external lead and hence has a very small decay rate falls off as a power law with the exponent $M - 1$. This result does not have a classical analogue.

\subsubsection{Large number of channels}

We can also consider the limit of a large number of channels $M \gg 1$ with perfect transparency $g = 1$. Applying saddle point approximation to the integral in Eq.\ (\ref{rhounitM}), we find
\begin{gather}
 \rho_{M \gg 1}(y)
  \approx \frac{M}{2y^2} f\left( \frac{y - M/2}{\sqrt{M}} \right), \label{rholargeM} \\
 f(z)
  = \frac{1}{2} \operatorname{erfc}(-\sqrt{2} z). \label{rhounitlargeM}
\end{gather}
At small values of $y$, the probability is negligibly small. At large $y$, the distribution follows the classical decay law $\sim M/(2y^2)$. Crossover between these two regions occurs at $y \approx M/2$ in the form of a relatively sharp step of the width $\propto \sqrt{M}$ and is described by the complementary error function \cite[\href{https://dlmf.nist.gov/7.2.E2}{(7.2.2)}]{DLMF}. We plot this crossover function below in Fig.\ \ref{Fig:crossover}.

The limiting distribution function of the form (\ref{rhounitlargeM}) is known in the theory of random non-Hermitian matrices \cite{Byun2025}. In the most general class of Gaussian complex matrices (Ginibre unitary ensemble, class A according to Ref.\ \cite{Kawabata2019}), eigenvalues are on average distributed uniformly inside a disk in the complex plane. At the edge of this disk, the density of the eigenvalues falls off to zero following the same error function law as in Eq.\ (\ref{rhounitlargeM}).

\subsubsection{Weak coupling limit}

Finally, we can study the limit of weak coupling $g \gg 1$. It corresponds to a relatively high barrier between the disordered system and the measuring probe. Hence the eigenenergies acquire only a small imaginary correction and scattering resonances remain well separated. We can simplify Eq.\ (\ref{rhounitM}) taking the limit $g \gg 1$ and obtain
\begin{equation}
 \rho_M(y)
  = \frac{g^M y^{M-1}}{(M - 1)!}\, e^{-gy}.
 \label{rhounitlargeg}
\end{equation}
This is the $\chi^2$ distribution with $2M$ degrees of freedom for the scaled variable $2gy$. This result is well-known in the context of random matrix theory as the generalized Porter-Thomas distribution \cite{Porter1956, Krieger1963, Alhassid1995}. It can be quite easily derived with the help of perturbation theory in the weak tunnel coupling \cite{Fyodorov2015}.

\subsection{Orthogonal class}

A disordered system with the time-reversal and spin symmetries is described by the sigma model of the orthogonal class. The matrix $Q$ has one compact fermionic Cartan variable $\lambda_\text{F}$ and two noncompact bosonic Cartan variables $\lambda_\text{B,B2}$. In the most general case, when the system is coupled to an external lead hosting $M$ channels with transparencies $g_n$, the function $\Gamma$ defined by Eqs.\ (\ref{SGamma}) and (\ref{Gamma}) takes the form
\begin{equation}
 \Gamma(\lambda)
  = \prod_{n = 1}^M \frac{g_n + \lambda_\text{F}}{\sqrt{g_n^2 + \lambda_\text{B}^2 + \lambda_\text{B2}^2 + 2 g_n \lambda_\text{B} \lambda_\text{B2} - 1}}.
  \label{Gammaorth}
\end{equation}
Analytic continuation in the variable $\lambda_\text{B}$ according to Eq.\ (\ref{tGammalambda}) provides a rather cumbersome expression. Each square root in the product (\ref{Gammaorth}) is imaginary in a certain domain of the two variables $\lambda_\text{B,B2}$ depending on the value of $g_n$. The whole product (\ref{Gammaorth}) is imaginary in the regions where some odd number of these domains overlap. A very general integral representation for $\rho(y)$ in a small metallic grain of orthogonal symmetry was given in Ref.\ \cite{Sommers1999}. It does conform to our general expression (\ref{rhoPsiGammalambda}) with the accurate accounting of all the relevant integration domains in $\lambda_\text{B,B2}$.

\subsubsection{General result for identical ideal channels}

We will not rederive this most general result here but instead consider a simpler case of $M$ identical channels all having perfect transparency $g = 1$. We construct a generating function with the auxiliary parameter $\alpha$ as in Eq.\ (\ref{Gammaalpha}) for this case
\begin{equation}
 \Gamma(\alpha, \lambda)
  = \frac{1 + \lambda_\text{F}}{\lambda_\text{B} + \lambda_\text{B2} + \alpha (1 + \lambda_\text{F})}.
\end{equation}
Performing analytic continuation (\ref{tGammalambda}) and taking imaginary part produces a delta function similar to the case of unitary class discussed above
\begin{equation}
 \tilde\Gamma(\alpha, \lambda)
  = 2(1 - \lambda_\text{F}) \delta\bigl[ \lambda_\text{B} - \lambda_\text{B2} - \alpha (1 - \lambda_\text{F}) \bigr].
\end{equation}
This delta function allows us to integrate over $\lambda_\text{B}$ and to recast Eq.\ (\ref{rhoPsiGammalambda}) as an integral in just two Cartan variables:
\begin{multline}
 \rho(\alpha, y)
  = \int \frac{d\lambda_\text{F}\, d\lambda_\text{B2}\, (1 + \lambda_\text{F})}
    {[2\lambda_\text{B2}^2 - \lambda_\text{F} - 1 + \alpha (1 - \lambda_\text{F})(2 \lambda_\text{B2} + \alpha)]^2} \\
      \times \frac{\partial^2 \Psi(\lambda)}{\partial y^2} \Bigr|_{\lambda_\text{B} = \lambda_\text{B2} + \alpha (1 - \lambda_\text{F})}.
 \label{rhoPsiorth}
\end{multline}
This is a general integral expression applicable to a disordered system of an arbitrary shape. In the simplest case of a small metallic grain, we can use Eq.\ (\ref{Psi0D}) that takes the following form in the orthogonal class:
\begin{equation}
 \Psi
  = e^{-y(\lambda_\text{B} \lambda_\text{B2} - \lambda_\text{F})}.
 \label{Psi0Dorth}
\end{equation}

With this function $\Psi$, we rewrite the integral in Eq.\ (\ref{rhoPsiorth}) in a relatively short form using the following notations:
\begin{subequations}
\begin{gather}
 t
  = \lambda_\text{B2}^2 - \lambda_\text{F} + \alpha \lambda_\text{B2} (1 - \lambda_\text{F}), \\
 d
  = 2t - (1 - \alpha^2)(1 - \lambda_\text{F}).
\end{gather}
\end{subequations}
Here $t$ is the function of $\lambda$ variables entering the exponent in Eq.\ (\ref{Psi0Dorth}) and $d$ is the denominator in the integrand of Eq.\ (\ref{rhoPsiorth}). Using these notations, we have
\begin{equation}
 \rho(\alpha, y)
  = \int_{-1}^1 d\lambda_\text{F} \int_1^\infty d\lambda_\text{B2}\, (1 + \lambda_\text{F}) \frac{t^2}{d^2}\, e^{-y t}.
 \label{rholambdaorth}
\end{equation}

There are many alternative ways to calculate this double integral. One obvious choice is to replace $\lambda_\text{F}$ with $t$ as an independent variable. Since the exponential factor depends only on $t$, integration over the other independent variable $\lambda_\text{B2}$ then becomes easy. However, subsequent integration over $t$ is quite tedious. It requires splitting the integrand in three terms and applying certain carefully chosen partial integrations to each of these terms.

We will present here a more straightforward path to the final result without making any changes of variables. Instead, we isolate a part of the integrand in Eq.\ (\ref{rholambdaorth}) that can be represented as a divergence of a two-dimensional vector in the $\lambda_\text{B2}$--$\lambda_\text{F}$ plane,
\begin{equation}
 \rho(\alpha, y)
  = \frac{1}{2} \int_{-1}^1 d\lambda_\text{F} \int_1^\infty d\lambda_\text{B2} \left[
      I - \frac{\partial I_\text{B2}}{\partial\lambda_\text{B2}} - \frac{\partial I_\text{F}}{\partial\lambda_\text{F}}
    \right].
 \label{rhoorthint}
\end{equation}
Three components of this expression are
\begin{subequations}
\begin{align}
 I_\text{B2}
  &= (1 - \lambda_\text{F}) (\lambda_\text{B2} + \alpha) \frac{t}{d}\, e^{-y t}, \\
 I_\text{F}
  &= (1 - \lambda_\text{F}^2) \frac{t}{d}\, e^{-y t}, \\
 I
  &= (1 - \lambda_\text{F}) \frac{\partial}{\partial y}\, y e^{-y t}.
\end{align}
\end{subequations}
We have thus absorbed all the complicated parts of the integrand into the derivative terms. The remaining bulk part $I$ is much easier to deal with since it has no denominator that couples integration variables.

The term with $I_\text{F}$ reduces to the surface contribution at $\lambda_\text{F} = \pm 1$ where it vanishes identically. The term with $I_\text{B2}$ yields a surface contribution at $\lambda_\text{B2} = 1$. It coincides with the $\lambda_\text{F}$ integral Eq.\ (\ref{rhounitint}) for the unitary class in the limit $g = 1$. Finally, in the bulk term $I$ the two integration variables are coupled only in the exponent and this coupling is proportional to $\alpha$. We can readily expand in powers of $\alpha$ up to the $(M-1)$th order and separate the two integrals. This yields the following explicit expression for the distribution function:
\begin{equation}
 \rho_M(y)
  = \left[1 + \frac{\partial}{\partial y}\frac{e^y\, \Gamma(M/2, y)}{2 y^{M/2 - 1}}\right]\frac{\gamma[M + 1, 2y]}{2y^2 (M - 1)!}.
 \label{rhoorthM}
\end{equation}
Derivative in $y$ here acts on everything to the right including the function outside the brackets. The result involves both lower \cite[\href{https://dlmf.nist.gov/8.2.E1}{(8.2.1)}]{DLMF} and upper \cite[\href{https://dlmf.nist.gov/8.2.E2}{(8.2.2)}]{DLMF} incomplete gamma functions.

\begin{figure}
    \includegraphics[width=0.47\textwidth]{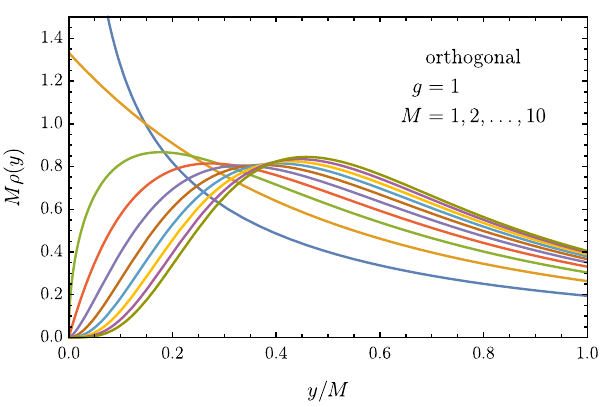}
    \caption{The density of scattering poles $\rho(y)$ in a small metallic grain of the orthogonal symmetry class, Eq.\ (\ref{rhoorthM}), attached to the measuring probe with $M$ perfectly conducting channels. Similar to Fig.\ \ref{Fig:resA}, we plot the distribution for the scaled variable $y/M$ to facilitate visual comparison of different curves.}
    \label{Fig:resAIg1}
\end{figure}

Equation (\ref{rhoorthM}) is one of our main findings. We have established an explicit closed form for the distribution function of scattering poles in the metallic grain of orthogonal symmetry attached to a perfect lead with $M$ conducting channels. For any odd $M$, this function can be further expressed in terms of the complementary error function \cite[\href{https://dlmf.nist.gov/7.2.E2}{(7.2.2)}]{DLMF}, while for even $M$ it reduces to elementary functions. As an example, the first two such distribution functions are
\begin{subequations}
\begin{align}
 \rho_{M=1}(y)
  &= -\frac{\partial}{\partial y} \biggl[
      \frac{1}{2y} - \frac{e^{-2y}}{2y} \notag \\
      \MoveEqLeft[-1] - \frac{\sqrt{\pi}}{4y^{3/2}} \bigl[ e^y - (1 + 2y) e^{-y} \bigr] \operatorname{erfc} \sqrt{y}
    \biggr], \\
 \rho_{M=2}(y)
  &= -\frac{\partial}{\partial y} \left[
      \frac{1}{y} - \frac{1}{2y^2} + \frac{e^{-2y}}{2y^2}
    \right].
\end{align}
\end{subequations}
Both these functions were first presented in Ref.\ \cite{Sommers1999} although the case $M=1$ was solved there in an integral form only. We plot the distribution Eq.\ (\ref{rhoorthM}) in Fig.\ \ref{Fig:resAIg1} for a few values of $M$.

\subsubsection{Asymptotics of large and small \texorpdfstring{$y$}{y}}

All the distribution functions Eq.\ (\ref{rhoorthM}) decay at large values of $y$. We can establish their asymptotic behavior either from Eq.\ (\ref{rhoorthM}) directly or from the integral (\ref{rholambdaorth}) expanding it in both $\lambda$ variables near $1$. Behavior at small $y$ is a power law which is most easily found from Eq.\ (\ref{rhoorthM}).
\begin{equation}
\rho_M(y)
 \approx \begin{dcases}
     \frac{M}{2 y^2} - \frac{M}{2 y^3} + \ldots, & y \gg 1, \\
     \frac{\sqrt{\pi} M y^{M/2 - 1}}{4\Gamma\bigl[ (M + 3)/2 \bigr]}, & y \ll 1.
   \end{dcases}
 \label{rhoorthasymp}
\end{equation}
Large $y$ limit reproduces the classical result (\ref{largeetatail}) much like it was in the unitary symmetry class, see Eq.\ (\ref{rhounitasymp}). However, unlike unitary symmetry, a negative subleading term appears in the large $y$ limit in the orthogonal class. This subleading term can be interpreted as a quantum correction to the classical result (\ref{largeetatail}), in some loose sense similar to the weak localization correction to conductivity \cite{Evers2008}. Small $y$ limit corresponds to the extreme quantum case hence the result is very much sensitive to the symmetry and has no classical interpretation.

\subsubsection{Large number of channels}

We can also estimate the distribution function for a large number of channels $M$. Similar to the unitary symmetry class, the distribution function has a relatively sharp step that connects a small value in the quantum limit with the classical decay at larger $y$. The crossover between these two limits can be quantified by using a uniform asymptotic expansion of the incomplete gamma functions \cite[\href{http://dlmf.nist.gov/8.12.E18}{(8.12.18)}]{DLMF} in Eq.\ (\ref{rhoorthM}) or from the integral (\ref{rhoorthint}) using saddle point approximation. The result has the same form (\ref{rholargeM}) as in the unitary class but with a different crossover function:
\begin{equation}
 f(z)
  = \frac{1}{2} \biggl[1 + \frac{\sqrt{\pi}}{4} \frac{\partial}{\partial z} e^{z^2} \operatorname{erfc}(z)\biggr] \operatorname{erfc}(-\sqrt{2} z).
\label{rhoorthlargeM}
\end{equation}
Derivative here acts on everything to the right similar to Eq.\ (\ref{rhoorthM}). We compare this crossover function to similar functions for other symmetry classes below in Fig.\ \ref{Fig:crossover}.

The distribution of the form (\ref{rhoorthlargeM}) was very recently found in Ref.\ \cite{Akemann2025} in the context of random non-Hermitian matrices. It was shown that the function (\ref{rhoorthlargeM}) describes the distribution of complex eigenvalues at the edge of the spectrum (disk in the complex plane) in the Gaussian ensemble of complex symmetric matrices (class AI$^\textit{\dag}$ according to Ref.\ \cite{Kawabata2019}). The fact that we observe the same distribution for the imaginary parts of the scattering poles demonstrates a great universality of such edge distributions.

\subsubsection{Weak coupling limit}

Consider the limit of tunnel junction when the measuring lead is coupled to the disordered system through a weak link with $g \gg 1$. In the case $g \neq 1$, it is more convenient to introduce new variables in the bosonic sector:
\begin{equation}
 \lambda_\pm
  = \lambda_\text{B} \lambda_\text{B2} \pm \sqrt{(\lambda_\text{B}^2 - 1)(\lambda_\text{B2}^2 - 1)}.
\end{equation}
These variables were used previously in Ref.\ \cite{Sommers1999}. They vary in the domain $\lambda_+ > \lambda_- > 1$ and for each pair of $\lambda_\pm$ there are two possible pairs of original variables up to the interchange $\lambda_\text{B} \leftrightarrow \lambda_\text{B2}$.

In terms of $\lambda_\pm$, the coupling function (\ref{tGammalambda}) takes the form
\begin{equation}
 \tilde\Gamma(\lambda)
  = -\frac{2}{\pi} \operatorname{Im} \frac{(g - \lambda_\text{F})^M}{(g - \lambda_- + i0)^{M/2} (g - \lambda_+ + i0)^{M/2}}.
\end{equation}
Imaginary part in this expression is nonzero only for $\lambda_+ \geq g$ and $1 \leq \lambda_- \leq g$. This allows us to simplify the integral in Eq.\ (\ref{rhoPsiGammalambda}) in the weak coupling case by assuming $\lambda_- \ll g$ and $\lambda_\text{F} \ll g$. The integration variables decouple in this limit and, after integrating over $\lambda_-$ and $\lambda_\text{F}$, we arrive at the following expression for the density of scattering resonances:
\begin{equation}
 \rho_M(y)
  = -\frac{1}{2\pi} \operatorname{Im}\int_g^\infty \frac{g^{M/2} e^{-y \lambda_+/2}\, d\lambda_+}{(g - \lambda_+ + i0)^{M/2}}.
\end{equation}

The remaining integral over $\lambda_+$ formally converges at the endpoint $\lambda_+ = g$ only for $M < 2$. For larger values of $M$ it should be understood as an integral over the Hankel contour that goes above the real axis from infinity to the point $\lambda_+ = g$, encircles this point counterclockwise, and returns back to infinity below the real axis. The difference of the integrand on the two halves of the contour at $\lambda_+ \pm i0$ will provide exactly the required imaginary part, while the whole integral remains finite and analytic in $M$. This kind of a contour integral is well known as one of the definitions of the gamma function \cite[\href{https://dlmf.nist.gov/5.9.E2}{(5.9.2)}]{DLMF}. With this result, the distribution of scattering resonances in the weak coupling limit becomes
\begin{equation}
 \rho_M(y)
  = \frac{g^{M/2} y^{M/2-1} e^{-gy/2}}{2^{M/2} \Gamma(M/2)}.
 \label{rhoorthlargeg}
\end{equation}

This is another version of the generalized Porter-Thomas distribution \cite{Porter1956, Krieger1963, Alhassid1995} in the form of the $\chi^2$ distribution with $M$ degrees of freedom for the scaled variable $gy$. Similarly to the case of unitary class, cf.\ Eq.\ (\ref{rhounitlargeg}), it can be derived directly from the random matrix theory for the disordered metallic grain perturbed by the weak tunnel coupling to the measuring probe \cite{Fyodorov2015}.

\subsection{Symplectic class}

Calculation of the scattering poles distribution in the symplectic symmetry class is quite similar to the orthogonal class considered above. The sigma model has only one noncompact bosonic variable $\lambda_\text{B} > 1$ and two compact fermionic variables $\lambda_\text{F,F2}$ with the constraint $-1 < \lambda_\text{F} < \lambda_\text{F2} < 1$, see Table \ref{Tab:classes}. There is one other important restriction for a system of symplectic symmetry: the number of channels $M$ in the measuring lead should be even and each transparency parameter $g_n$ is doubly degenerate. This degeneracy is a direct consequence of the Kramers theorem due to the broken spin symmetry.

\subsubsection{General result for identical ideal channels}

Generic form of the function $\Gamma$ defined by Eqs.\ (\ref{SGamma}) and (\ref{Gamma}) is
\begin{equation}
 \Gamma(\lambda)
  = \prod_{n = 1}^{M/2} \frac{g_n^2 + \lambda_\text{F}^2 + \lambda_\text{F2}^2 + 2 g_n \lambda_\text{F} \lambda_\text{F2} - 1}{(g_n + \lambda_\text{B})^2}.
 \label{Gammasymp}
\end{equation}
As we did previously in the orthogonal class, we will restrict our consideration only to the case of $M$ identical channels with perfect transparency $g = 1$. We will define a generating function with the parameter $\alpha$ for arbitrary values of $M$ in the standard way:
\begin{multline}
 \Gamma(\alpha, \lambda)
  = \sum_{M = 1}^\infty (-\alpha)^{M-1} \Gamma_M(\lambda) \\
  = \frac{\lambda_\text{F} + \lambda_\text{F2}}{1 + \lambda_\text{B} + \alpha(\lambda_\text{F} + \lambda_\text{F2})}.
\end{multline}
This function includes all positive integer $M$, both even and odd, while only even numbers of channels have physical significance in the symplectic class. We nevertheless find subsequent calculations easier with this extended generating function. Meaningful results can be drawn in the end by taking only odd powers (corresponding to even $M$) in the small $\alpha$ expansion.

Analytic continuation of $\Gamma$ according to Eq.\ (\ref{tGammalambda}) yields
\begin{equation}
 \tilde\Gamma(\lambda) 
  = (\lambda_\text{F2} - \lambda_\text{F}) \delta[1 - \lambda_\text{B} - \alpha(\lambda_\text{F2} - \lambda_\text{F})].
\end{equation}
We substitute this $\tilde\Gamma$ in the integral Eq.\ (\ref{rhoPsiGammalambda}) for the density of poles and reduce it to the double integral over compact $\lambda$ variables only taking advantage of the delta function. For a symplectic system of arbitrary geometry with the known function $\Psi$, this leads to the general expression
\begin{multline}
 \rho(\alpha, y)
  = \int \frac{d\lambda_\text{F}\, d\lambda_\text{F2}\, \alpha [2 + \alpha(\lambda_\text{F2} - \lambda_\text{F})]}
    {2[(1 + \alpha^2)(\lambda_\text{F2} - \lambda_\text{F}) + 2 \alpha (1 - \lambda_\text{F2} \lambda_\text{F})]^2} \\
      \times \frac{\partial^2 \Psi(\lambda)}{\partial y^2} \Bigr|_{\lambda_\text{B} = 1 + \alpha(\lambda_\text{F2} - \lambda_\text{F})}.
 \label{rhoPsisymp}
\end{multline}

As it was done for other symmetry classes above, we will limit our consideration to the case of a small metallic grain where Eq.\ (\ref{Psi0D}) applies and yields
\begin{equation}
 \Psi
  = e^{-y(\lambda_\text{B}  - \lambda_\text{F2} \lambda_\text{F})}.
\end{equation}
We insert this function in Eq.\ (\ref{rhoPsisymp}) and use the notations
\begin{subequations}
\begin{gather}
 t
  = 1 + \alpha (\lambda_\text{F2} - \lambda_\text{F}) - \lambda_\text{F2} \lambda_\text{F},\\
 d
  = (1 + \alpha^2)(\lambda_\text{F2} - \lambda_\text{F}) + 2 \alpha (1 - \lambda_\text{F2} \lambda_\text{F})
\end{gather}
\end{subequations}
for the exponent and denominator of the integrand. The density of scattering resonances then takes the form
\begin{equation}
 \rho(\alpha, y)
  = \frac{\alpha}{2} \int_{-1}^1 d\lambda_\text{F} \int_{\lambda_\text{F}}^1 d\lambda_\text{F2} [2 + \alpha(\lambda_\text{F2} - \lambda_\text{F})] \frac{t^2}{d^2} e^{-yt}.
 \label{rholambdasymp}
\end{equation}

As it was already discussed in the case of orthogonal class, there are several alternative approaches to calculating this integral. The most straightforward method is to apply a change of variables by making $\lambda_\text{F2}$ and $t$ the new independent variables. Since the exponential factor depends only on $t$, integration over $\lambda_\text{F2}$ can be carried out first. Then the remaining integral over $t$ splits into several parts that require a careful analysis. Instead, we will provide here a more direct route to the final result by isolating in the integrand of Eq.\ (\ref{rholambdasymp}) a total divergence of a vector in the $\lambda_\text{F}$--$\lambda_\text{F2}$ plane. 
\begin{equation}
 \rho(\alpha, y)
  = \frac{1}{4} \int_{-1}^1 d\lambda_\text{F} \int_{\lambda_\text{F}}^1 d\lambda_\text{F2} \left[
      \frac{\partial I_\text{F}}{\partial\lambda_\text{F}} + \frac{\partial I_\text{F2}}{\partial\lambda_\text{F2}} - I
    \right].
 \label{rhosympJ}
\end{equation}
Components of this expression are
\begin{subequations}
\begin{align}
 I_\text{F}
  &= (\lambda_\text{F2} - \lambda_\text{F}) (\alpha \lambda_\text{F} - 1) \frac{t}{d}\, e^{-y t}, \\
 I_\text{F2}
  &= (\lambda_\text{F2} - \lambda_\text{F}) (\alpha \lambda_\text{F2} + 1) \frac{t}{d}\, e^{-y t}, \\
 I
  &= (\lambda_\text{F2} - \lambda_\text{F}) \frac{\partial}{\partial y}\, y e^{-y t}.
\end{align}
\end{subequations}

Integration in Eq.\ (\ref{rhosympJ}) runs over a triangular region. The vector $\{I_\text{F}, I_\text{F2}\}$ vanishes at the diagonal $\lambda_\text{F} = \lambda_\text{F2}$. Hence the integral of its divergence reduces to the flow of this vector through the two sides of the triangle at $\lambda_\text{F} = -1$ and $\lambda_\text{F2} = 1$. This part of the integral equals the distribution function (\ref{rhounitint}) in the unitary class. In the remaining two-dimensional integral of $I$ the variables almost decouple in terms of $\lambda_\text{F} \pm \lambda_\text{F2}$. Namely, the integrand $I$ indeed splits into a product of functions of these two combinations but the shape of the integration domain does not allow to separate the variables completely. For this reason, we cannot derive a closed expression for the density or scattering resonances for a given value of $M$ as it was done in the orthogonal symmetry class. Instead, we can carry out integration for a fixed value of $\alpha$. We also take the odd part of $\rho$ in $\alpha$ thus keeping only the contributions of physically relevant even values of $M$ and obtain the result
\begin{multline}
 \rho(\alpha, y)
  = \frac{\alpha}{4} \frac{\partial}{\partial y} \biggl[
      \frac{1 - e^{-2 (1 + \alpha) y}}{(1 + \alpha)^2 y} \\
      + e^{-(1 - \alpha ^2) y} \Bigl( \operatorname{Ei}\bigl[ (1-\alpha^2) y \bigr]
      - \operatorname{Ei}\bigl[ -(1 + \alpha)^2 y \bigr] \Bigr)
   \\ + \{ \alpha \mapsto -\alpha \} \biggr].
 \label{rhosympA}
\end{multline}
This function is written in term of the exponential integral \cite[\href{https://dlmf.nist.gov/6.2.E5}{(6.2.5)}]{DLMF}.

\begin{figure}
    \includegraphics[width=0.47\textwidth]{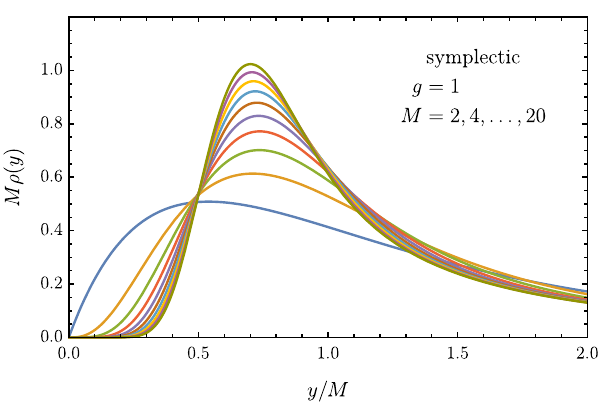}
    \caption{The density of scattering poles $\rho(y)$ in a small metallic grain of the symplectic symmetry class [Eq.\ (\ref{rhosympA}) expanded in a power series in $\alpha$] attached to the measuring probe with $M$ perfectly conducting channels. Similar to Figs.\ \ref{Fig:resA} and \ref{Fig:resAIg1}, we plot the distribution for the scaled variable $y/M$ to facilitate visual comparison of different curves.}
    \label{Fig:resAIIg1}
\end{figure}

Expanding Eq.\ (\ref{rhosympA}) in a power series in $\alpha$ produces distribution functions for individual values of $M$. Here are the first two such distributions:
\begin{subequations}
\begin{align}
 \rho_{M = 2}(y)
  &= -\frac{\partial}{\partial y} \left[
      \frac{1}{2y} - \frac{e^{-2y}}{2y}
      + e^{-y} \operatorname{Shi} y
    \right], \label{symplrhoM2} \\
 \rho_{M = 4}(y)
  &= -\frac{\partial}{\partial y} \left[
      \frac{3}{2y} - \left( \frac{3}{2y} + \frac{3}{2} \right) e^{-2y}
      + y e^{-y} \operatorname{Shi} y
    \right].
\end{align}
\end{subequations}
They involve the hyperbolic sine integral \cite[\href{https://dlmf.nist.gov/6.2.E15}{(6.2.15)}]{DLMF}. We show these and a few more of the distribution functions in Fig.\ \ref{Fig:resAIIg1}.

\subsubsection{Asymptotics of large and small \texorpdfstring{$y$}{y}}

Similar to other symmetry classes, the distribution function of scattering resonances in the symplectic case has a power-law asymptotics at small values of $y$ (quantum limit) and decays at large $y$ (classical limit). Both asymptotics can be extracted from the integral (\ref{rhosympJ}). At large values of $y$, the integral is determined by the vicinity of the point $\lambda_\text{F} = \lambda_\text{F2} = 1$. For $y \ll 1$, we can first expand the integrand in powers of $\alpha$ up to the $(M-1)$th order and then expand the remaining exponential factor in small $y$. The resulting asymptotics are
\begin{equation}
\rho_M(y)
 \approx \begin{dcases}
     \frac{M}{2 y^2} + \frac{M}{2 y^3} + \ldots, & y \gg 1, \\
     \frac{2^{M + 1} M y^{M - 1}}{(M + 2)!}, & y \ll 1.
   \end{dcases}
 \label{rhosympasymp}
\end{equation}

We observe here the same universal classical tail (\ref{largeetatail}) at large $y$ as we have already seen in other classes, Eqs.\ (\ref{rhounitasymp}) and (\ref{rhoorthasymp}). The positive subleading term in the limit $y \gg 1$ represents a small quantum correction to the classical result (\ref{largeetatail}). This term differs only by its sign from the similar term in the orthogonal class (\ref{rhoorthasymp}) and is in some sense similar to the weak antilocalization correction that also differs only by the overall sign from the weak localization in the orthogonal class \cite{Evers2008}. Power-law decay at small $y$ corresponds to the extreme quantum limit of the problem and does not have a classical analogue.

\subsubsection{Large number of channels}

When the number of channels $M$ is large, the distribution function is strongly suppressed at small $y$. Similar to other symmetry classes, it undergoes a sharp crossover from an almost zero value to the classical decay pattern at larger $y$. This crossover happens at $y \sim M/2$ and has a characteristic width $\propto \sqrt{M}$. In order to derive the crossover function, we expand $I$ in the integrand of Eq.\ (\ref{rhosympJ}) to the $(M-1)$th power in small $\alpha$ and then estimate the result by applying saddle point approximation for $M \gg 1$. This leads to the universal dependence (\ref{rholargeM}) with the crossover function
\begin{equation}
 f(z)
  = \int_{-z}^\infty \frac{dx}{\sqrt{8\pi}}\, e^{-(x - z)^2/2} \left[ 1 + \sqrt{\pi}\, z\, e^{x^2} \operatorname{erfc} x \right].
 \label{rhosymplargeM}
\end{equation}
Due to the lack of variable separation in the integral (\ref{rhosympJ}), this crossover function is also represented in an integral form only.

Previously, we saw that the crossover functions (\ref{rhounitlargeM}) and (\ref{rhoorthlargeM}) coincide with edge distributions for the complex eigenvalues of the general complex Gaussian matrices and symmetric complex Gaussian matrices, respectively. From this analogy, we conjecture that the function (\ref{rhosymplargeM}) should coincide with the edge distribution of the eigenvalues for the complex Gaussian ensemble of symmetric quaternion matrices (class AII$^\textit{\dag}$ according to Ref.\ \cite{Kawabata2019}). The latter is any matrix $H$ that obeys the constraint $H = s^y H^T s^y$ with the second Pauli matrix $s^y$ in the spin space. As far as we are aware, such an edge distribution in class AII$^\textit{\dag}$ is as yet unknown.

\begin{figure}
    \includegraphics[width=0.47\textwidth]{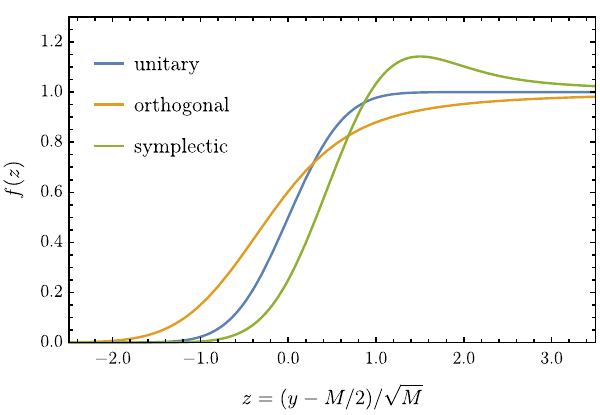}
    \caption{Crossover functions $f(z)$ describing the distribution of scattering poles (\ref{rholargeM}) in the limit $M \gg 1$ for all three Wigner-Dyson classes, Eqs.\ (\ref{rhounitlargeM}), (\ref{rhoorthlargeM}), and (\ref{rhosymplargeM}).}
    \label{Fig:crossover}
\end{figure}

We compare the three crossover functions (\ref{rhounitlargeM}), (\ref{rhoorthlargeM}), and (\ref{rhosymplargeM}) for the three symmetry classes in Fig.\ \ref{Fig:crossover}. All three functions interpolate between $0$ at large negative arguments and $1$ at large positive arguments. At large values of $z$ they behave as
\begin{equation}
 f(z \gg 1)
  \approx \begin{dcases}
    1, &\text{unitary}, \\
    1 - \frac{1}{4z^2}, &\text{orthogonal}, \\
    1 + \frac{1}{4z^2}, &\text{symplectic}.
  \end{dcases}
\end{equation}
These asymptotics qualitatively show the onset of the negative and positive quantum correction in the orthogonal and symplectic symmetry class, respectively, cf.\ the limit $y \gg 1$ in Eqs.\ (\ref{rhoorthasymp}) and (\ref{rhosympasymp}). Let us note, however, that these corrections at large $z$ are not accurate quantitatively. This discrepancy occurs because the accuracy of the crossover approximation near $y \sim M/2$ is insufficient to reproduce correctly the subleading terms in the limit $y \gg M/2$.

\subsubsection{Weak coupling limit}

Consider now the limit of weak coupling between the measuring lead and the disordered metallic grain, $g \gg 1$. In the function $\Gamma$, Eq.\ (\ref{Gammasymp}), we can neglect both $\lambda_\text{F}$ and $\lambda_\text{F2}$ in comparison with $g$. Analytic continuation (\ref{tGammalambda}) then yields a high derivative of the delta function:
\begin{equation}
 \tilde\Gamma
  = -\frac{1}{\pi} \operatorname{Im} \frac{g^M}{(g - \lambda_\text{B} + i0)^M}
  = \frac{g^M\,  \delta^{(M-1)}(\lambda_\text{B} - g)}{(M-1)!}.
\end{equation}
We can also neglect $\lambda_\text{F,F2}$ in the integral (\ref{rhoPsiGammalambda}) since $\lambda_\text{B} \sim g \gg 1$. The resulting density of resonances is then
\begin{equation}
 \rho_M(y)
  = \int_1^\infty d\lambda_\text{B}\, e^{-y\lambda_\text{B}} \tilde\Gamma(\lambda_\text{B})
  = \frac{g^M y^{M-1}}{(M - 1)!}\, e^{-gy}.
 \label{rhosymplargeg}
\end{equation}
This result coincides identically with the similar expression (\ref{rhounitlargeg}) in the unitary symmetry class but is valid for even values of $M$ only. It is the same $\chi^2$ distribution with $2M$ degrees of freedom for the scaled variable $2gy$. As was already discussed before, this is an instance of the generalized Porter-Thomas distribution \cite{Porter1956, Krieger1963, Alhassid1995} that can be derived from the spectrum of a random matrix perturbed by a weak tunnel coupling to the measuring lead \cite{Fyodorov2015}.

\subsubsection{Single pair of channels}

The structure of the function $\Gamma$ in the symplectic class, Eq.\ (\ref{Gammasymp}), allows us to solve analytically the case of a single pair of channels, $M = 2$, with an arbitrary coupling $g$. Analytic continuation (\ref{tGammalambda}) of this function yields a derivative of the delta function in the bosonic variable:
\begin{equation}
 \tilde\Gamma
  = (g^2 + \lambda_\text{F}^2 + \lambda_\text{F2}^2 - 2 g \lambda_\text{F} \lambda_\text{F2} - 1)\, \delta'(\lambda_\text{B} - g).
\end{equation}
This makes it possible to integrate over $\lambda_\text{B}$ in Eq.\ (\ref{rhoPsiGammalambda}) and express the density of resonances as a double integral in $\lambda_\text{F,F2}$. We use the notations
\begin{subequations}
\begin{gather}
 t
  = g - \lambda_\text{F} \lambda_\text{F2}, \\
 d
  = g^2 + \lambda_\text{F}^2 + \lambda_\text{F2}^2 - 2 g \lambda_\text{F} \lambda_\text{F2} - 1.
\end{gather}
\end{subequations}
and represent this integral as
\begin{equation}
 \rho(y)
  = \int_{-1}^1 d\lambda_\text{F,F2}\, \frac{\partial^2}{\partial y^2} \left[ \frac{(g^2 - 1)(4t + yd)}{4d^2} - \frac{g}{2d} \right] e^{-yt}.
\end{equation}
The integrand is fully symmetric with respect to the interchange of variables $\lambda_\text{F} \leftrightarrow \lambda_\text{F2}$ hence we can extend the integration domain to the full square $-1 < \lambda_\text{F,F2} < 1$.

\begin{figure}
    \includegraphics[width=0.47\textwidth]{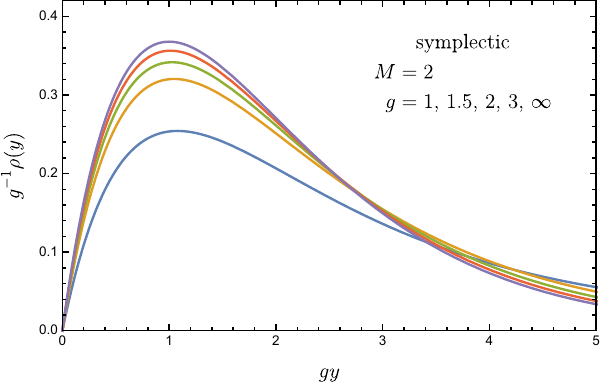}
    \caption{The density of scattering resonances $\rho(y)$ in a small metallic grain of symplectic symmetry attached to the measuring probe with $M = 2$ conducting channels, Eq.\ (\ref{rhosympM2}). The distribution is shown for the scaled variable $gy$ to demonstrate a smooth crossover from Eq.\ (\ref{symplrhoM2}) for $g = 1$ to Eq.\ (\ref{rhosymplargeg}) in the limit $g \gg 1$.}
    \label{Fig:resAIIM2}
\end{figure}

As it was already done on several occasions previously, we can isolate a total divergence of a vector in the integrand and thus get rid of the denominator that couples the variables
\begin{equation}
 \rho(y)
  = \frac{1}{4} \int_{-1}^1 d\lambda_\text{F,F2} \left[
      \frac{\partial I_\text{F}}{\partial\lambda_\text{F}} + \frac{\partial I_\text{F2}}{\partial\lambda_\text{F2}} + I
    \right].
\end{equation}
Three terms in this expression contain
\begin{subequations}
\begin{align}
 I_\text{F}
  &= (g \lambda_\text{F} - \lambda_\text{F2}) \frac{t^2}{d}\, e^{-yt}, \\
 I_\text{F2}
  &= (g \lambda_\text{F2} - \lambda_\text{F}) \frac{t^2}{d}\, e^{-yt}, \\
 I
  &= \frac{\partial^2}{\partial y^2}\, y e^{-yt}.
\end{align}
\end{subequations}
Integration of the divergence part provides a surface contribution from the edges of the square $-1 < \lambda_\text{F,F2} < 1$. The bulk part with $I$ can be integrated directly due to the absence of the denominator. The resulting density of scattering poles is
\begin{equation}
 \rho(y)
  = -\frac{\partial}{\partial y}\, e^{-gy} \left( \frac{\sinh y}{y} + g \operatorname{Shi} y \right).
 \label{rhosympM2}
\end{equation}
This function is shown in Fig.\ \ref{Fig:resAIIM2} for several values of $g$. It interpolates smoothly between the result (\ref{symplrhoM2}) for the ideal lead with $g = 1$ and the weak coupling limit Eq.\ (\ref{rhosymplargeg}) with $M = 2$ in the limit $g \gg 1$.

\section{Numerical study}
\label{Sec:numerics}

In the previous Section, we have presented a number of results for the density of scattering resonances for small metallic grains of the unitary, orthogonal, and symplectic symmetry classes. All these results can be directly compared to the numerical simulations of random matrices in these classes. We will carry out such a comparison in this Section.

A small metallic grain of a given symmetry is modeled by a random Hamiltonian $H$ in the form of an $N \times N$ matrix \cite{Gorkov1965} whose elements are independent and obey the Gaussian distribution with the zero mean and the following second order moment:
\begin{equation}
 \langle H_{ij} H_{kl} \rangle
  = 4N \times \begin{dcases}
      \delta_{il} \delta_{kj} , &\text{unitary}, \\
      \delta_{il} \delta_{kj} + \delta_{ik} \delta_{lj} , &\text{orthogonal}, \\
      \delta_{il} \delta_{kj} + s^y_{ik} s^y_{lj} , &\text{symplectic}.
    \end{dcases}
 \label{RMT}
\end{equation}
In the unitary case, the Hamiltonian is a generic Hermitian matrix, orthogonal class corresponds to real symmetric matrices, and symplectic class is represented by the symmetric matrices composed of real quaternions \cite{Mehta2004}. In the latter case, each element is itself a $2 \times 2$ matrix and $s^y$ is the corresponding second Pauli matrix in the spin space.

Average density of states in these Gaussian ensembles has a semicircle shape \cite{Mehta2004}. We have chosen the prefactor $4N$ in Eq.\ (\ref{RMT}) such that the mean level spacing in the middle of the semicircle band is $\Delta = 2\pi$. Then the eigenvalues of the Hamiltonian are automatically normalized according to the convention (\ref{y}). The semicircle itself occupies the window $[-4N,\,+4N]$ that gets larger with increasing the matrix size.

Attaching the measuring probe to the metallic grain modeled by the ramdom-matrix Hamiltonian is equivalent to adding negative imaginary parts to the diagonal elements of the matrix:
\begin{equation}
 H_\text{eff}
  = H - 2i N \operatorname{diag} \left\{ g_n - \sqrt{g_n^2 - 1} \right\}.
 \label{Heidelberg}
\end{equation}
This effective non-Hermitian Hamiltonian appears naturally in the framework of the so-called Heidelberg approach \cite{Verbaarschot1985} to  quantum chaotic scattering. It has been actively discussed as a starting point for addressing the statistics of the scattering matrix resonances in Refs.\ \cite{Sokolov1989, Fyodorov1997}. The number of nonzero entries in the diagonal matrix added in Eq.\ (\ref{Heidelberg}) equals the number $M$ of channels in the measuring lead. In the symplectic case, each channel is doubly degenerate due to the Kramers theorem hence the added matrix automatically contains $M/2$ entries each proportional to the unit $2 \times 2$ matrix.

\begin{figure*}
\begin{center}
    \includegraphics[width = 0.47 \textwidth]{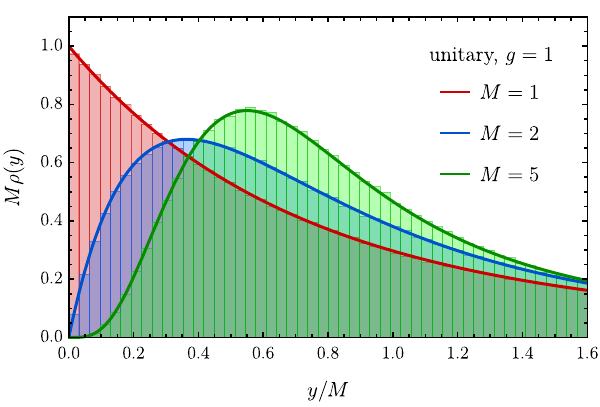}\hfill\includegraphics[width = 0.47 \textwidth]{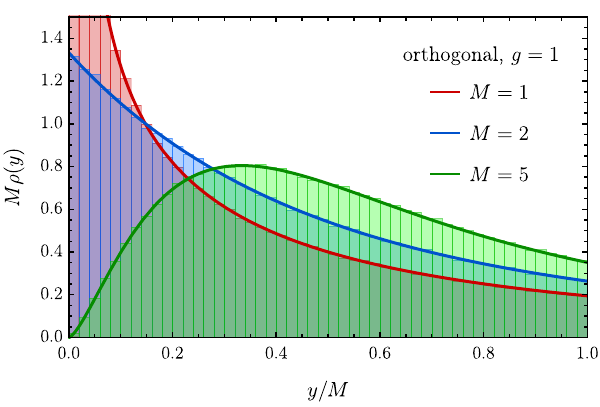}\\
    \includegraphics[width = 0.47 \textwidth]{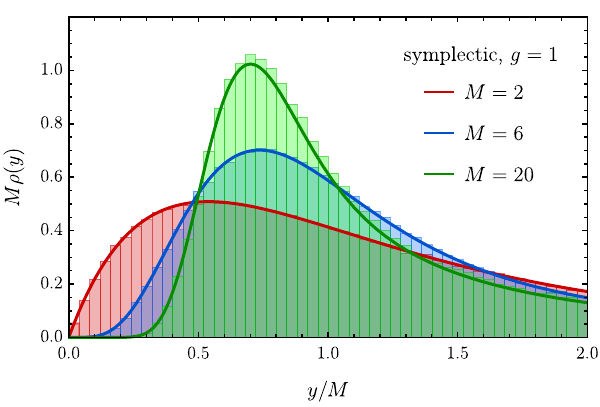}\hfill\includegraphics[width = 0.47 \textwidth]{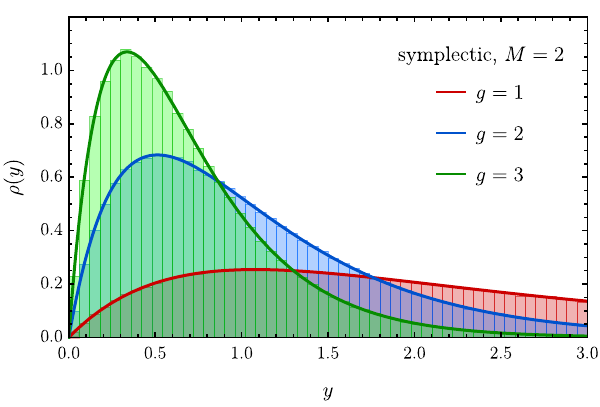}
\end{center}
\caption{Distributions of the imaginary parts of the eigenvalues of the effective Hamiltonian (\ref{Heidelberg}). Solid curves show the corresponding analytical results. Upper left panel: unitary class with perfect transparency $g = 1$ compared to Eq.\ (\ref{rhounitM}). Upper right panel: orthogonal class with $g = 1$ and Eq.\ (\ref{rhoorthM}). Lower left panel: symplectic class with $g = 1$ and Eq.\ (\ref{rhosympA}) (after expansion in small $\alpha$ to the appropriate order). Lower right panel: symplectic class with a single pair of channels $M = 2$ and Eq.\ (\ref{rhosympM2}).}
\label{Fig:numerics}
\end{figure*}

For each set of parameters, we generate random effective Hamiltonians (\ref{Heidelberg}) of the size $N = 1000$ and perform their numerical diagonalization. Then $10\%$ of the eigenvalues with the smallest absolute values (in the middle of the spectrum) are selected. This is done in order to mitigate the curvature of the semicircle spectrum. The procedure is repeated $10^4$ times to collect the necessary statistics. Histograms of the imaginary parts of these eigenvalues are presented in Fig.\ \ref{Fig:numerics} together with the corresponding analytical expressions from the previous Section. We observe a perfect agreement between our theory and numerical simulations in all the studied cases.

\section{Summary and discussion}
\label{Sec:summary}

To summarize, we have developed a very general approach to calculating the average distribution of scattering resonances in disordered systems. Our method is based on the supersymmetric nonlinear sigma model that includes on equal footing both the disordered system itself and the clean measuring lead attached to it. Such an approach is enabled thanks to the specifically designed source terms in the sigma-model action inside the probe. Our general result is contained in Eqs.\ (\ref{rhoF}), (\ref{FZ}), and (\ref{ZQ}).

Based on this very general expression for the distribution of scattering poles, we have derived the classical Moldauer-Simonius \cite{Moldauer1967, Simonius1974} relation (\ref{aveta}) for the average decay rate in a disordered system of any symmetry and geometry coupled to the measuring lead through a generic potential barrier.

We have further performed a Fourier analysis of the convolution integral (\ref{ZQ}) on the sigma-model manifold and have reduced it to the integral over Cartan parameters of the matrix $Q$ only. This reduction dramatically simplifies the calculation of the distribution function. In the most general form, the result is given by Eqs.\ (\ref{tGamma}) and (\ref{rhoPsiGamma}) in terms of $\theta$ variables or, alternatively, by Eqs.\ (\ref{tGammalambda}) and (\ref{rhoPsiGammalambda}) in terms of $\lambda$ variables.

We have applied these general equations to the case of small disordered metallic grains, where the sigma-model partition function reduces to a very simple expression (\ref{Psi0D}). This allowed us to reproduce all the known results \cite{Fyodorov1997} for the unitary symmetry class, significantly extend the results of Ref.\ \cite{Sommers1999} in the orthogonal class, and derive previously unknown distribution functions in the symplectic class. We have found analytic expressions for the distribution of scattering poles in all three symmetry classes for a given number of channels $M$ with perfect coupling $g = 1$ and analyzed various asymptotic forms of these functions. Our findings are summarized in Table \ref{Tab:results} with references to specific equations and figures.

\begin{table}
\begin{center}
\begin{tblr}{ccc}
\hline\hline
Unitary & Orthogonal & Symplectic \\
\hline
\SetCell[c=3]{c}{General result: $M$ channels with same $g$} \\
   {Eq.\ (\ref{rhounitM}), Fig.\ \ref{Fig:resA} \\ {[any $M$, any $g$]}}
 & {Eq.\ (\ref{rhoorthM}), Fig.\ \ref{Fig:resAIg1} \\ {[any $M$, $g = 1$]}}
 & {Eq.\ (\ref{rhosympA})\hyperlink{foot}{$^a$}, Fig.\ \ref{Fig:resAIIg1} \\ {[even $M$, $g = 1$]}} \\
&& {Eq.\ (\ref{rhosympM2}), Fig.\ \ref{Fig:resAIIM2} \\ {[$M = 2$, any $g$]}} \\
\hline
\SetCell[c=3]{c}{Asymptotics $y \gg 1$ and $y \ll 1$ [any $M$, $g = 1$]:} \\
Eq.\ (\ref{rhounitasymp}) & Eq.\ (\ref{rhoorthasymp}) & Eq.\ (\ref{rhosympasymp}) \\
\hline
\SetCell[c=3]{c}{Thick lead [$M \gg 1$, $g = 1$]: Fig.\ \ref{Fig:crossover}, Eq.\ (\ref{rholargeM}) with}\\
Eq.\ (\ref{rhounitlargeM}) & Eq.\ (\ref{rhoorthlargeM}) & Eq.\ (\ref{rhosymplargeM}) \\
\hline
\SetCell[c=3]{c}{Weak coupling limit [any $M$, $g \gg 1$]:}\\
Eq.\ (\ref{rhounitlargeg}) & Eq.\ (\ref{rhoorthlargeg}) & Eq.\ (\ref{rhosymplargeg}) \\
\hline\hline
\SetCell[c=3]{l}{\hypertarget{foot}{\footnotesize $^a$ As a generating function of $\alpha$.}}
\end{tblr}
\end{center}
\caption{Summary of results for the distribution function of scattering poles in small metallic grains of different symmetries. The table refers to specific equations and figures in Sec.\ \ref{Sec:0D}.}
\label{Tab:results}
\end{table}

Our general method can be used to find the distribution function of scattering poles in several other cases. First, it is quite straightforward to apply Eqs.\ (\ref{tGammalambda}) and (\ref{rhoPsiGammalambda}) to the semi-infinite disordered wires of orthogonal and symplectic symmetry similar to how it was done in Ref.\ \cite{Fyodorov2024} for the unitary class. The sigma-model partition function (zero mode of the transfer-matrix Hamiltonian) in these cases is already known \cite{Khalaf2017} hence it only remains to compute the corresponding integrals. Second, the same method can be used to find quantum corrections to the classical distribution of scattering poles (limit $y \gg 1$) in systems of any symmetry and geometry. It amounts to developing a perturbative expansion of the function $\Psi$ in inverse $y$ and then applying the general equations derived in the present paper. Third, it seems plausible that the general result, that universally applies to the three standard Wigner-Dyson symmetry classes, can be generalized to describe the crossover between these classes. Such a situation arises in the case when time-reversal symmetry is only weakly violated and is especially relevant for most experimental setups used in recent measurements. Finally, it is also worth exploring possible applications of our approach to other unconventional symmetry classes, e.g., in superconducting samples. Extending our research in all these directions will be the goal of further publications.

\acknowledgments{We are very much indebted to Yan Fyodorov for bringing our attention to scattering resonances in disordered systems and for numerous illuminating discussions of this topic. We are also grateful to M.\ Feigel'man, D.\ Ivanov, A.\ Levin, A.\ Lunkin, and I.\ Poboiko for discussions and valuable advices. We would like to express our special gratitude to S.\ Gureva for providing computing resources.}

\appendix

\section{Generating function in terms of Green's functions}
\label{App:Nazarov}

In this Appendix, we derive Eq.\ (\ref{FcheckG}) for the generating function (\ref{F}) in terms of the matrix Green function (\ref{checkG}). We begin with specifying the Hamiltonian inside the lead at $x > 0$, see Fig.\ \ref{Fig:system}, and discussing ordinary retarded and advanced Green functions.

The measuring probe is a fully clean ballistic one-dimensional conductor with a given set of propagating modes. We will use the basis of these conducting modes and denote the right- and left-propagating wave functions as $\ket{+_n}$ and $\ket{-_n}$, respectively. They are normalized to the unit particle flow:
\begin{equation}
 \braket{+_n}{+_m} = \braket{-_n}{-_m} = \frac{\delta_{m n}}{v_n},
 \quad
 \braket{-_n}{+_m} = 0.
\end{equation}
Here $v_n$ is the group velocity of the $n$th propagating mode. Complete dependence of the wave function on the transverse coordinates as well as possible fast longitudinal oscillations with the Fermi wave length are included in $\ket{\pm}$. The effective Hamiltonian in this basis acts only on relatively slow envelope amplitudes of the wave function in different channels and hence can be written in the linearized form
\begin{equation}
 H = - i \hat v \frac{\partial}{\partial x},
 \quad
 \hat v = \sum_n \bigl( \ket{+_n} v_n^2 \bra{+_n} - \ket{-_n} v_n^2 \bra{-_n} \bigr).
 \label{Hv}
\end{equation}

At the point $x = 0$, the lead is attached to the disordered sample. Full scattering (reflection) matrix $S_E$ of the sample imposes boundary conditions on the wave functions at $x = 0$. Namely, an eigenstate of the Hamiltonian with the eigenvalue $E$ describing a particle incident in the channel $\ket{-_n}$ and reflected as a linear combination of the $\ket{+_n}$ states has the form
\begin{equation}
 \ket{\psi_n}
  = \ket{-_n} e^{-iEx/v_n} + \sum_m \ket{+_m} e^{iEx/v_m} S_{mn}.
 \label{psin}
\end{equation}
Any eigenstate of the Hamiltonian that obeys boundary conditions can be represented as a linear combination of these $\ket{\psi_n}$ functions.

Retarded and advanced Green functions are solutions to the following defining equation:
\begin{align}
 \left( E + i \hat v \frac{\partial}{\partial x} \pm i0 \right) G^{R/A}_E (x_1, x_2)
  = \delta(x_1 - x_2).
 \label{Green}
\end{align}
For $x_1 < x_2$, Green functions conform to the boundary conditions and hence their $x_1$ dependence is given by a linear combination of the scattering states (\ref{psin}). For $x_1 > x_2$, retarded/advanced Green function should contain only right-/left-moving components in $x_1$, respectively. This follows from the condition that the Green function does not grow at infinity and from the presence of the $\pm i0$ term added to the energy. Finally, at the point $x_1 = x_2$, Green functions have a jump due to the delta function in Eq.\ (\ref{Green}). Together, all these requirements fully determine both Green functions:
\begin{subequations}
\label{GreenRA}
\begin{align}
 G^R_E(x_1, x_2)
  &= -i\sum_{n,m} \ket{+_n} e^{iEx_1/v_n} S_{nm} e^{iEx_2/v_m} \bra{-_m} \notag \\
  \MoveEqLeft[3.2] -i\sum_n \begin{cases}
      \ket{+_n} e^{iE(x_1 - x_2)/v_n} \bra{+_n}, & x_1 > x_2, \\
      \ket{-_n} e^{-iE(x_1 - x_2)/v_n} \bra{-_n}, & x_1 < x_2,
    \end{cases} \\
 G^A_E (x_1, x_2)
  &= i\sum_{n,m} \ket{-_n} e^{-iEx_1/v_n} S^{-1}_{nm} e^{-iEx_2/v_m} \bra{+_m} \notag \\
  \MoveEqLeft[3.2] +i\sum_n \begin{cases}
      \ket{-_n} e^{-iE(x_1 - x_2)/v_n} \bra{-_n}, & x_1 > x_2, \\
      \ket{+_n} e^{iE(x_1 - x_2)/v_n} \bra{+_n}, & x_1 < x_2.
    \end{cases}
\end{align}
\end{subequations}

Using the explicit Green functions and taking the limit $x_{1,2} \to +0$, we have the following relation:
\begin{multline}
 \lim_{x_{1,2} \to +0} \operatorname{tr} \bigl[ \hat v G^A_{E - i\eta}(x_1, x_2) \hat v G^{R}_{E + i\eta}(x_2, x_1) \bigr]^n \\
  = \operatorname{tr} \bigl[1 - S_{E - i\eta}^{-1} S_{E + i\eta} \bigr]^n.
 \label{Kubo}
\end{multline}
Note that this identity is equally valid when the limit is taken either for $x_1 > x_2$ or for $x_1 < x_2$ although individual Green functions are different in these two cases. It is also valid for any integer value of $n > 0$. In deriving Eq.\ (\ref{Kubo}), we have also taken advantage of the analyticity of the Green functions when shifting the energy into upper/lower complex half-plane by $i\eta$ for the retarded/advanced function. 

Now we have all the ingredients to evaluate the determinant in Eq.\ (\ref{FcheckG}). As for any block matrix, it can be rewritten in terms of individual blocks, i.e., retarded and advanced Green functions and velocity operators. Up to an unimportant constant factor (independent of $a$), we have 
\begin{multline}
 \ln \det \check G \\
  = \operatorname{tr} \ln \Bigl[ 1 - \sin^2(a/2) \hat v G^{A}_{E - i \eta}(x_1, x_2) \hat v G^{R}_{E + i \eta}(x_2, x_1) \Bigr].
\label{checkGdet}
\end{multline}
We expand the logarithm in the right-hand side in the standard power series, take the limit $x_{1,2} \to 0$, and apply the identity (\ref{Kubo}) to each term of this expansion. Summing the series back together, we have the result
\begin{multline}
 \ln \det \check G(x_{1,2} = 0) \\
  = \operatorname{tr} \ln \Bigl[ \cos^2(a/2) + \sin^2(a/2) S_{E - i\eta}^{-1} S_{E + i\eta} \Bigr].
\end{multline}
Equation (\ref{FcheckG}) is exactly this identity at $a = \pi$.

\section{Derivation of the sigma model with source terms}
\label{App:sigma}

In this Appendix, we derive the nonlinear sigma-model action in the presence of sources terms $a_\text{B,F}$. Our derivation follows the ideas of Refs.\ \cite{Rejaei1996, Khalaf2016}. We will show that the presence of source parameters fixes the value $Q = Q_a$ inside the lead, see Eq.\ (\ref{Qa}). To avoid technical complications, we will consider here the model of the unitary class only. The derivation can be easily adapted to the orthogonal and symplectic classes as well, see Ref.\ \cite{Khalaf2016}.

The supersymmetric partition function of any noninteracting system (\ref{Z}) can be written as a Gaussian integral over a four-dimensional complex superfield $\phi$ that contains two commuting and two Grassmann components and thus belongs to the direct product of RA and BF spaces. The partition function has the form
\begin{gather}
    Z(a_\text{B,F}) = \int D\phi^*\, D\phi\, e^{-S[\phi^*, \phi]}, \\
    S
     = S_\text{samp}
    -i\int_0^\infty dx\, \phi^\dag \Lambda \Biggl[ E + i\hat v \frac{\partial}{\partial x} - V_\text{dis} + i \eta \Lambda
    \notag \\ 
    +
    \hat v \sin(\hat{a}/2) \begin{pmatrix} 0 & \delta(x - x_2) \\ \delta(x - x_1) & 0 \end{pmatrix}
    \Biggr] \phi. 
\label{Sphi}
\end{gather}
The first part of the action, $S_\text{samp}$, describes the disordered sample. We focus here instead on the second part associated with the measuring lead at $x > 0$, see Fig.\ \ref{Fig:system}. This is done because the source fields are present only there.

The Hamiltonian in the lead has the standard one-dimensional ballistic form (\ref{Hv}). We also add a random potential term $V_\text{dis}$ to this Hamiltonian in order to facilitate the derivation of the sigma model. In the end, the strength of disorder inside the measuring lead will be taken to zero. Diagonal matrix $\hat a$ was defined in Eq.\ (\ref{Qa}). Gaussian integrals over commuting and anticommuting parts of the field yield the denominator and the numerator of the partition function (\ref{Z}), respectively.

We apply a gauge transformation to the fields $\phi$ and $\phi^\dag$ in order to get rid of the source terms:
\begin{equation}
 \phi \mapsto U(x) \phi, \qquad \phi^\dag \mapsto \phi^\dag \Lambda U^{-1}(x) \Lambda.
\end{equation}
This is achieved if the matrix $U(x)$ satisfies the equation
\begin{equation}
 \frac{dU(x)}{dx} = i \sin(\hat{a}/2) \begin{pmatrix} 0 & \delta(x - x_2) \\ \delta(x - x_1) & 0 \end{pmatrix} U(x).
\end{equation}
Suppose for definiteness $x_1 < x_2$. Then a general solution to the above equation is
\begin{gather}
  U =
    \begin{cases}
        U_0, & x < x_1,\\
        \begin{pmatrix} 1 & 0 \\ i \sin(\hat{a}/2) & 1 \end{pmatrix} U_0, & x_1 < x < x_2,\\
        \begin{pmatrix} \cos^2(\hat{a}/2) &  i \sin(\hat{a}/2) \\ i \sin(\hat{a}/2) & 1 \end{pmatrix} U_0, & x > x_2.
    \end{cases}
\end{gather}
For now, $U_0$ is an arbitrary constant matrix to be specified later. After the gauge transformation, the action (\ref{Sphi}) takes the form
\begin{equation}
    S = S_\text{samp} - i \int_0^\infty dx\, \phi^\dag \Lambda \bigl(E - H + i \eta\, U^{-1} \Lambda U \bigr) \phi.
\end{equation}
The sources are now hidden in the $\eta$ term.

We choose the matrix $U_0$ of the form
\begin{equation}
 U_0
  = \begin{pmatrix} 1 & 0 \\ 0 & i \cos(\hat a/2) \end{pmatrix}.
\end{equation}
First, it is diagonal and commutes with $\Lambda$. Hence, at $x < x_1$ the action remains completely unchanged. At the same time, for $x > x_2$ we obtain a balanced matrix that coincides with $Q_a$, cf.\ Eq.\ (\ref{Qa}):
\begin{equation}
    U^{-1} \Lambda U =
    \begin{cases}
        \Lambda, & x < x_1,\\
        Q_a, & x > x_2.
    \end{cases}
\end{equation}
We can now take the limit $x_{1,2} = 0$ and proceed with the standard derivation of the sigma model \cite{Efetov1996}.

The resulting action in the disordered part of the system (sample) will have the form of Eq.\ (\ref{sigma}). It is fully independent of source parameters. The part of the action in the measuring lead will be also the standard one-dimensional sigma model but with the altered matrix $\Lambda$:
\begin{equation}
  S_\text{lead}[Q]
   = \frac{\pi \nu}{4} \int_0^\infty dx\, \operatorname{str} \left[D \left( \frac{\partial Q}{\partial x} \right)^2 - 4 \eta Q_a Q \right].
\end{equation}
This action appears due to the disorder potential $V_\text{dis}$ that was previously introduced inside the measuring lead, cf.\ Eq.\ (\ref{Sphi}). We will now gradually reduce this disorder potential in order to restore ballistic transport there. This will increase the value of the diffusion constant $D$ in the lead and make gradients of $Q$ very unfavorable. We thus conclude that $Q$ remains an $x$-independent constant in the limit of a ballistic lead. The value of this constant is determined by the term with $\eta$ and equals $Q_a$.

We have thus proven that the source terms fix the value of $Q = Q_a$ inside the measuring lead. This proof relies on the sigma-model action inside the lead. Hence, strictly speaking, it is only valid in the limit $M \gg 1$. The conclusion is however more general. Fixing $Q = Q_a$ in the lead is actually correct for any $M$, but would require a more sophisticated derivation in the general case.

\section{Fourier analysis on the sigma-model manifolds}
\label{App:eigenfunctions}

In this Appendix, we discuss main properties and provide explicit forms of the eigenfunctions of the radial Laplace-Beltrami operator (\ref{Delta}) on the sigma-model manifolds. More details on this topic can be found in Refs.\ \cite{Mirlin1994, KhalafDisser}.

\subsection{Unitary class}

The sigma model of the unitary class involves two Cartan angles $\theta_\text{B,F}$, see Table \ref{Tab:classes}. To make equations more compact, we will mainly use the alternative $\lambda$ variables, see Eq.\ (\ref{lambda}).

The Jacobian of the Cartan parametrization has the following forms in terms of these variables:
\begin{equation}
 J(\theta)
  = \frac{\sinh\theta_\text{B} \sin\theta_\text{F}}{(\cosh\theta_\text{B} - \cos\theta_\text{B})^2}, \quad
 J(\lambda)
  = \frac{1}{(\lambda_\text{B} - \lambda_\text{F})^2}.
\end{equation}
In terms of $\lambda$ variables, the radial part of the Laplace-Beltrami operator (\ref{Delta}) can be written as \cite{Efetov1996}
\begin{equation}
 \Delta
  = \frac{1}{J} \left[
      \frac{\partial}{\partial\lambda_\text{B}} J (\lambda_\text{B}^2 - 1) \frac{\partial}{\partial\lambda_\text{B}}
      + \frac{\partial}{\partial\lambda_\text{F}} J (1 - \lambda_\text{F}^2) \frac{\partial}{\partial\lambda_\text{F}}
    \right].
\end{equation}

Eigenfunctions of this operator (zonal spherical functions) are given explicitly by the products of Legendre functions:
\begin{equation}
 L_\mathbf{q}(\lambda)
  = -\frac{(l + 1/2)^2 + q^2}{2} (\lambda_\text{B} - \lambda_\text{F}) P_{-1/2 + iq}(\lambda_\text{B}) P_l(\lambda_\text{F}).
 \label{Lunit}
\end{equation}
Here the momentum multi-index $\mathbf{q}$ stands for the pair of one discrete parameter $l$ and one continuous parameter $q$, see Table \ref{Tab:classes}. The corresponding eigenvalue of the Laplace-Beltrami operator is
\begin{equation}
 \Delta L_\mathbf{q}(\lambda)
  = -\bigl[ (l + 1/2)^2 + q^2 \bigr] L_\mathbf{q}(\lambda).
\end{equation}

The eigenfunctions (\ref{Lunit}) are properly normalized to satisfy the identities (\ref{normalization}). Summation/integration measure for the momentum $\mathbf{q}$ is defined with the weight function $\mu_\mathbf{q}$:
\begin{subequations}
\begin{gather}
 \sum_\mathbf{q} \ldots
  = \sum_{l = 0}^\infty \int_0^\infty dq\, \mu_\mathbf{q} \ldots, \\
 \mu_\mathbf{q}
  = \frac{2 (2l + 1) q \tanh(\pi q)}{[(l + 1/2)^2 + q^2]^2}.
\end{gather}
\end{subequations}

\subsection{Orthogonal class}

In the orthogonal class, there are three Cartan variables: two noncompact bosonic angles $\theta_\text{B,B2} > 0$ and one compact fermionic angle $0 < \theta_\text{F} < \pi$, see Table \ref{Tab:classes}. The Jacobian of this parametrization is
\begin{multline}
 J(\theta)
  = \sinh\theta_\text{B} \sinh\theta_\text{B2} \sin^3\theta_\text{F} \\
  \times [\cosh(\theta_\text{B} + \theta_\text{B2}) - \cos\theta_\text{F}]^{-2} \\
  \times [\cosh(\theta_\text{B} - \theta_\text{B2}) - \cos\theta_\text{F}]^{-2}.
\end{multline}
In terms of $\lambda$ variables, Eq.\ (\ref{lambda}), it takes a more concise form
\begin{equation}
 J(\lambda)
  = \frac{1 - \lambda_\text{F}^2}{(\lambda_\text{F}^2 + \lambda_\text{B}^2 + \lambda_\text{B2}^2 - 2 \lambda_\text{F} \lambda_\text{B} \lambda_\text{B2} - 1)^2}.
\end{equation}
With this Jacobian, the radial part of the Laplace-Beltrami operator takes the form \cite{Efetov1996}
\begin{multline}
 \Delta
  = \frac{1}{J} \biggl[
      \frac{\partial}{\partial\lambda_\text{F}} J (1 - \lambda_\text{F}^2) \frac{\partial}{\partial\lambda_\text{F}}
      + \frac{\partial}{\partial\lambda_\text{B}} J (\lambda_\text{B}^2 - 1) \frac{\partial}{\partial\lambda_\text{B}} \\
      + \frac{\partial}{\partial\lambda_\text{B2}} J (\lambda_\text{B2}^2 - 1) \frac{\partial}{\partial\lambda_\text{B2}}
    \biggr].
 \label{Delta_orth}
\end{multline}

Eigenfunctions of this operator are labeled by one discrete (integer) parameter $l \geq 0$ and two continuous parameters $q, q_2 > 0$. We combine them into a single momentum index $\mathbf{q}$. In order to present eigenfunctions explicitly, we introduce several auxiliary notations. First, define the function
\begin{equation}
 R_\nu^\mu(\lambda)
  = \begin{dcases}
      (\lambda^2 - 1)^{\mu/2} P_\nu^\mu(\lambda), & \lambda > 1, \\ 
      (1 - \lambda^2)^{\mu/2} \mathsf{P}_\nu^\mu(\lambda), & -1 < \lambda < 1.
    \end{dcases}
 \label{R}
\end{equation}
This is a generalization of the associated Legendre function that works equally well for compact and noncompact variables. Here $P_\nu^\mu(\lambda)$ is the standard associated Legendre function \cite[\href{https://dlmf.nist.gov/14.3.E6}{(14.3.6)}]{DLMF} and $\mathsf{P}_\nu^\mu(\lambda)$ is the so called associated Legendre function ``on the cut'' also known as the Ferrers function \cite[\href{https://dlmf.nist.gov/14.3.E1}{(14.3.1)}]{DLMF}. Second, we introduce a notation for the product of three $R$ functions
\begin{equation}
 R^{abc}_\mathbf{q}(\lambda)
  = R_l^a(\lambda_\text{F}) R_{-1/2 + iq}^b(\lambda_\text{B}) R_{-1/2 + iq_2}^c(\lambda_\text{B2}).
\end{equation}
These three functions have their degrees fixed by the components of $\mathbf{q}$. Third, we will make use of the short-hand notations
\begin{equation}
 F = l(l + 1), \quad B = q^2 + \frac{1}{4}, \quad  B_2 = q_2^2 + \frac{1}{4}.
\end{equation}

With all the above definitions, we have the following general eigenfunction of the operator (\ref{Delta_orth}):
\begin{align}
 L_\mathbf{q}(\lambda)
  \MoveEqLeft[1] = R^{011}_\mathbf{q}(\lambda) \notag \\ 
    &+ \frac{F + B + B_2}{2} (\lambda_\text{F} - \lambda_\text{B} \lambda_\text{B2}) R^{000}_\mathbf{q}(\lambda) \notag \\
    &+ \frac{F + B - B_2}{2} (\lambda_\text{B} - \lambda_\text{F} \lambda_\text{B2}) R^{-110}_\mathbf{q}(\lambda) \notag \\
    &+ \frac{F - B + B_2}{2} (\lambda_\text{B2} - \lambda_\text{F} \lambda_\text{B}) R^{-101}_\mathbf{q}(\lambda) \notag \\
    &+ \left[ \frac{(F + B + B_2)^2}{4} - B B_2 \right] \notag \\
    \MoveEqLeft[-1] \times (\lambda_\text{F}^2 + \lambda_\text{B}^2 + \lambda_\text{B2}^2 - 2 \lambda_\text{F} \lambda_\text{B} \lambda_\text{B2} - 1) R^{-100}_\mathbf{q}(\lambda).
 \label{Lorth}
\end{align}
This function was derived in Ref.\ \cite{KhalafDisser} with the help of Iwasawa parametrization of the sigma-model manifold. It is properly normalized to satisfy identities (\ref{normalization}). The eigenvalue is
\begin{equation}
 \Delta L_\mathbf{q}(\lambda)
  = - \bigl( F + B + B_2 \bigr) L_\mathbf{q}(\lambda).
\end{equation}

Summation over all values of $\mathbf{q}$ is defined with the weight function $\mu_\mathbf{q}$ as
\begin{subequations}
\begin{gather}
 \sum_\mathbf{q} \ldots
  = \sum_{l = 0}^\infty \int_0^\infty dq\, dq_2\, \mu_\mathbf{q} \ldots, \\
 \mu_\mathbf{q}
  = 8 l (l + 1) (2l + 1) q q_2 \tanh(\pi q) \tanh(\pi q_2) \notag \\
    \times \bigl[ l^2 + (q + q_2)^2 \bigr]^{-1} \bigl[ (l + 1)^2 + (q + q_2)^2 \bigr]^{-1} \notag \\
    \times \bigl[ l^2 + (q - q_2)^2 \bigr]^{-1} \bigl[ (l + 1)^2 + (q - q_2)^2 \bigr]^{-1}.
\end{gather}
\end{subequations}
This weight function was first found in Ref.\ \cite{Mirlin1994}. It formally vanishes for $l = 0$ and should be understood as the limit $l \to +0$. In this limit, the weight $\mu_\mathbf{q}$ acquires an extra delta function
\begin{equation}
 \mu_{l=0,q,q_2}
  = \frac{\tanh^2(\pi q)}{2\pi (q^2 + 1/4)} \delta(q - q_2).
\end{equation}
This means that the eigenfunctions with $l = 0$ are only normalizable for $q = q_2$ and can be indexed with a single continuous parameter $q$.

\subsection{Symplectic class}

The sigma model of the symplectic class is in many respects similar to the case of orthogonal symmetry. Spherical functions on the sigma-model manifold in these two classes are actually related to each other by analytic continuation. There are two compact fermionic Cartan angles $0 < \theta_\text{F2} < \theta_\text{F} < \pi$ and one noncompact bosonic angle $\theta_\text{B} > 0$, see Table\ \ref{Tab:classes}. Jacobian of the Cartan parametrization is
\begin{multline}
 J(\theta)
  = \sinh^3\theta_\text{B} \sin\theta_\text{F} \sin\theta_\text{F2} \\
  \times [\cosh\theta_\text{B} - \cos(\theta_\text{F} + \theta_\text{F2})]^{-2} \\
  \times [\cosh\theta_\text{B} - \cos(\theta_\text{F} - \theta_\text{F2})]^{-2}.
\end{multline}
In terms of $\lambda$ variables, Eq.\ (\ref{lambda}), it becomes
\begin{equation}
 J(\lambda)
  = \frac{\lambda_\text{B}^2 - 1}{(\lambda_\text{B}^2 + \lambda_\text{F}^2 + \lambda_\text{F2}^2 - 2 \lambda_\text{B} \lambda_\text{F} \lambda_\text{F2} - 1)^2}.
\end{equation}
Radial part of the Laplace-Beltrami operator (\ref{Delta}) has the following form in these variables:
\begin{multline}
 \Delta
  = \frac{1}{J} \biggl[
      \frac{\partial}{\partial\lambda_\text{B}} J (\lambda_\text{B}^2 - 1) \frac{\partial}{\partial\lambda_\text{B}}
      + \frac{\partial}{\partial\lambda_\text{F}} J (1 - \lambda_\text{F}^2) \frac{\partial}{\partial\lambda_\text{F}} \\
      + \frac{\partial}{\partial\lambda_\text{F2}} J (1 - \lambda_\text{F2}^2) \frac{\partial}{\partial\lambda_\text{F2}}
    \biggr].
\end{multline}

A generic eigenfunction is labeled by two integer parameters $l, l_2 > 0$ that have the same parity and with one continuous parameter $q > 0$. These three quantities are combined in a single momentum index $\mathbf{q}$. Similar to the case of orthogonal class, we will use the functions (\ref{R}) and define the triple product
\begin{equation}
 R^{abc}_\mathbf{q}(\lambda)
  = R_{-1/2 + iq}^a(\lambda_\text{B}) R_l^b(\lambda_\text{F}) R_{l_2}^c(\lambda_\text{F2})
\end{equation}
and the short-hand notations
\begin{equation}
 B = q^2 + \frac{1}{4}, \quad F = l(l + 1), \quad F_2 = l_2(l_2 + 1).
\end{equation}
With these definitions, the general spherical function can be written as
\begin{align}
 L_\mathbf{q}(\lambda) \MoveEqLeft[1] = R^{011}_\mathbf{q}(\lambda) \notag \\ 
 &- \frac{B + F + F_2}{2} (\lambda_\text{B} - \lambda_\text{F} \lambda_\text{F2}) R^{000}_\mathbf{q}(\lambda) \notag \\
 &- \frac{B + F - F_2}{2} (\lambda_\text{F} - \lambda_\text{B} \lambda_\text{F2}) R^{-110}_\mathbf{q}(\lambda) \notag \\
 &- \frac{B - F + F_2}{2} (\lambda_\text{F2} - \lambda_\text{B} \lambda_\text{F}) R^{-101}_\mathbf{q}(\lambda) \notag \\
 &+ \left[ \frac{(B + F + F_2)^2}{4} - F F_2 \right] \notag \\
 \MoveEqLeft[-1] \times (\lambda_\text{B}^2 + \lambda_\text{F}^2 + \lambda_\text{F2}^2 - 2 \lambda_\text{B} \lambda_\text{F} \lambda_\text{F2} - 1) R^{-100}_\mathbf{q}(\lambda).
\end{align}
This function was first derived in Ref.\ \cite{KhalafDisser} with the help of Iwasawa parametrization. Similar to all previous cases, it is properly normalized to satisfy identities (\ref{normalization}). The eigenvalue is
\begin{equation}
\Delta L_\mathbf{q}(\lambda)
 = - \bigl( B + F + F_2 \bigr) L_\mathbf{q}(\lambda).
\end{equation}

Summation over all values of $\mathbf{q}$ is defined with the proper weight function:
\begin{subequations}
\begin{gather}
 \sum_\mathbf{q} \ldots
  = \sum_{\substack{l, l_2 \geq 0 \\ \text{$l + l_2$ even}}} \int_0^\infty dq\, \mu_\mathbf{q} \ldots, \\
 \mu_\mathbf{q}
  = 8 (2 l + 1) (2 l_2 + 1) (q^2 + 1/4) q \tanh(\pi q) \notag \\
    \times \bigl[ (l + l_2 + 3/2)^2 + q^2 \bigr]^{-1} \bigl[ (l + l_2 + 1/2)^2 + q^2 \bigr]^{-1} \notag \\
    \times \bigl[ (l - l_2 + 1/2)^2 + q^2 \bigr]^{-1}  \bigl[ (l - l_2 - 1/2)^2 + q^2 \bigr]^{-1}.
\end{gather}
\end{subequations}
This weight function was first found in Ref.\ \cite{Mirlin1994}. Unlike the case of orthogonal class, there are no any specific values of $\mathbf{q}$ that would require special treatment.

\bibliography{resonances}

\end{document}